\documentclass[aps,pra,twocolumn,superscriptaddress,nofootinbib]{revtex4-2}

\usepackage[utf8]{inputenc}
\usepackage{amsmath}    
\usepackage{amssymb}    
\usepackage{graphicx}   
\usepackage{bm}         
\usepackage[english]{babel}

\usepackage{physics}  \usepackage[colorlinks=true,allcolors=blue]{hyperref}

\begin{document}


\title{Stochastic phase-space simulation of multimode cat states via the positive-$P$ representation}

\author{Yi Shi (Corresponding author)}
\thanks{ucapysh@ucl.ac.uk}
\affiliation{Department of Physics and Astronomy, University College London, Gower Street, London WC1E 6BT, United Kingdom}

\author{Alex Ferrier}
\affiliation{Department of Physics and Astronomy, University College London, Gower Street, London WC1E 6BT, United Kingdom}
\affiliation{Center for Theoretical Physics, Polish Academy of Sciences, Aleja Lotnik\'ow 32/46, 02-668 Warsaw, Poland}

\author{Piotr Deuar}
\affiliation{Institute of Physics, Polish Academy of Sciences,
Aleja Lotników 32/46, PL-02-668 Warsaw, Poland}

\author{Eran Ginossar}
\affiliation{School of Mathematics and Physics,
University of Surrey, Guildford GU2 7XH, United Kingdom}

\author{Marzena Szymanska}
\affiliation{Department of Physics and Astronomy, University College London, Gower Street, London WC1E 6BT, United Kingdom}
\affiliation{Center for Theoretical Physics, Polish Academy of Sciences, Aleja Lotnik\'ow 32/46, 02-668 Warsaw, Poland}

\date{\today}

\begin{abstract}
\begin{center}
\textbf{Abstract} \\
\end{center}
We present a comprehensive study of the transient dynamics of multimode Schrödinger cat states in dissipatively coupled resonator arrays. Using the positive-$P$ phase-space method, we derive the exact stochastic differential equations governing the system's dynamics, enabling the simulation of system sizes significantly larger than those accessible via direct master equation simulation. We demonstrate the utility of this method by simulating transient dynamics for networks up to $N=21$ sites. Furthermore, we critically examine the method's usefulness and limitations, specifically highlighting the computational instability encountered when estimating the state parity in the systems. Our results provide a pathway for scalable simulations of non-Gaussian states in large open quantum systems. 
\end{abstract}

\maketitle


\section{Introduction}

Schrödinger cat states, macroscopic quantum superpositions of coherent states, have emerged as a cornerstone resource in the pursuit of fault-tolerant quantum computation \cite{mirrahimi2014dynamically,grimm2020stabilization}. By encoding quantum information into the infinite-dimensional Hilbert space of a harmonic oscillator, these bosonic codes offer a hardware-efficient alternative to conventional multi-qubit registers \cite{cai2021bosonic}. A defining advantage of cat-state encodings is their intrinsic noise bias. With appropriate stabilization, bit-flip errors can be exponentially suppressed, leaving only phase-flip errors to be corrected by a simple classical repetition code \cite{guillaud2019repetition,puri2020bias}. Recently, this paradigm has been extended from single-mode cavities to multimode architectures. In particular, Zapletal {\it et al.}~\cite{zapletal2022stabilization} proposed a scalable scheme for stabilizing multimode Schrödinger cat states in arrays of coupled resonators via normal-mode dissipation engineering. Their work suggests that distributing the cat state across multiple modes yields a noise bias that scales exponentially with the system size, offering a path toward highly robust quantum memories.

However, the theoretical validation of such multimode protocols faces a significant computational bottleneck. The Hilbert space dimension of a coupled resonator array grows exponentially with the number of modes $N$, rendering exact numerical solutions of the Lindblad master equation intractable for all but the smallest systems ($N = 3$ in \cite{zapletal2022stabilization}). While analytic solutions for the steady-state manifold exist, understanding the transient dynamics requires numerical treatments that lie beyond the reach of standard density matrix simulations. To characterize the dynamics of realistic large-scale implementations, efficient simulation techniques capable of handling high-dimensional open quantum systems are required.
\begin{figure}
    \centering
    \includegraphics[width=\linewidth]{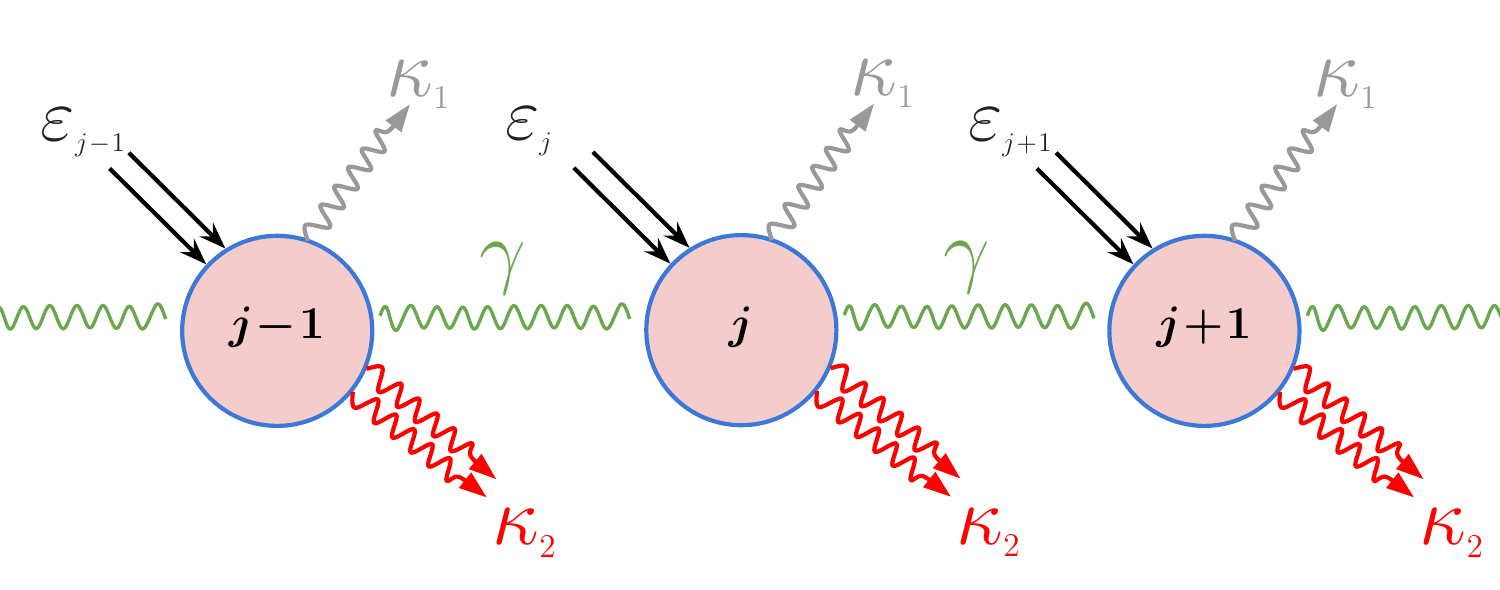}
    \caption{Schematic of the driven-dissipative resonator array: a linear-cavity chain with engineered non-local dissipative coupling of strength \(\gamma\), parametric two-photon drive of amplitude \(\epsilon\), and local single- and two-photon losses at rates \(\kappa_1\) and \(\kappa_2\), respectively.}
    \label{fig:fig_0}
\end{figure}

\begin{figure}
    \centering
    \includegraphics[width=\linewidth]{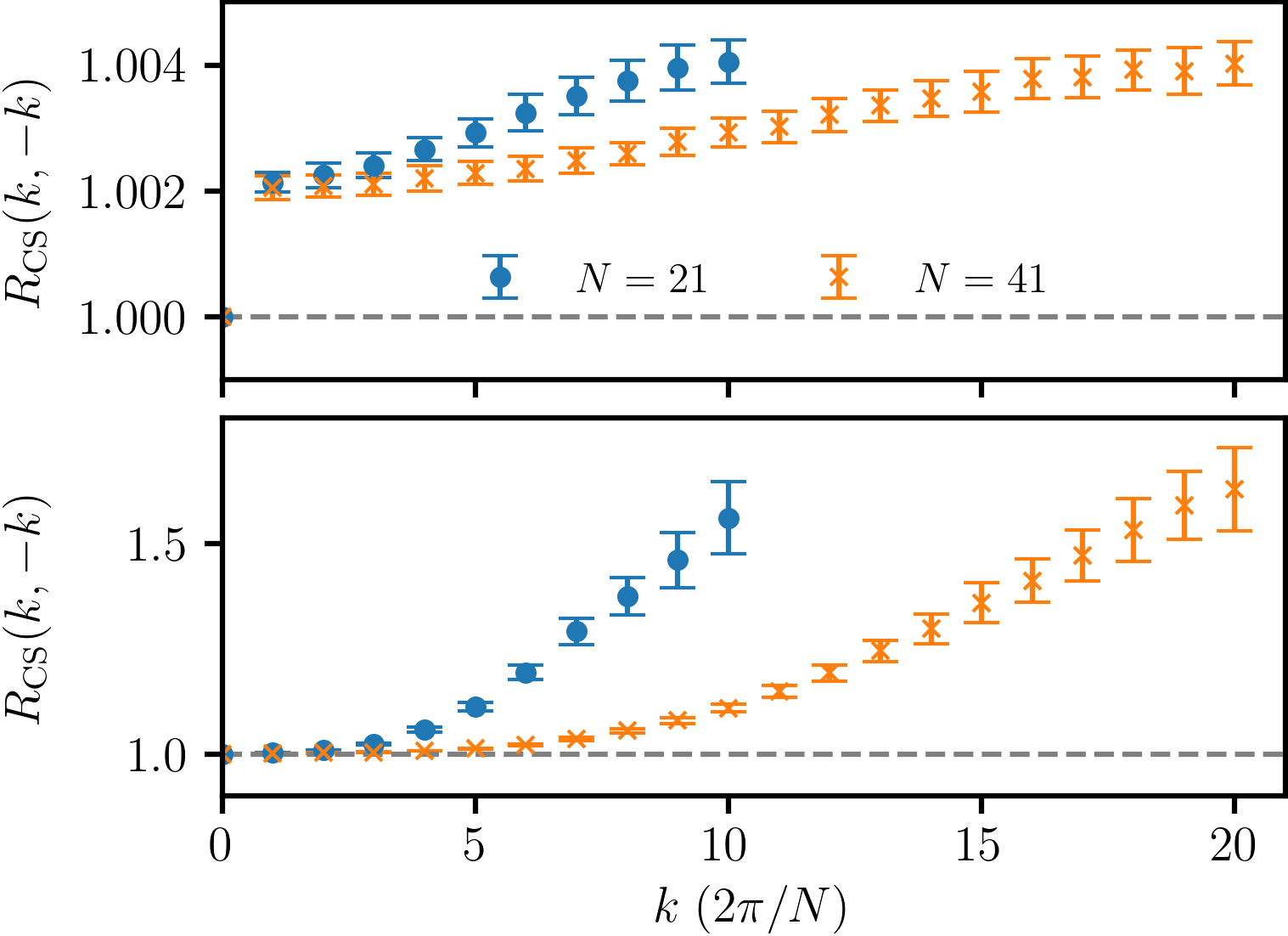}
    \caption{
    Nonclassical correlations in dissipatively coupled resonator arrays.
    Persistent-state normally ordered Cauchy--Schwarz ratio
    $R_{\mathrm{CS}}(k,-k)$ between opposite-quasimomentum
    modes for $N=21$ and $N=41$ resonators at
    (a) $\gamma/\epsilon=0.2$ and
    (b) $\gamma/\epsilon=2$, where $\gamma$ is the
    dissipative coupling strength and $\epsilon$ is the
    two-photon drive amplitude.
    Values above the classical bound
    $R_{\mathrm{CS}}=1$ (dashed line) witness nonclassical
    correlations. Error bars indicate standard errors
    estimated from stochastic subensembles.
    The correlation measure and detailed analysis are
    presented in Sec.~\ref{sec:Quasimomentum space}.
    }
    \label{fig:m7}
\end{figure}

In this work, we address this scalability challenge by applying the positive-$P$ stochastic simulation method \cite{gilchrist1997positive,deuar2021fully} to the linear resonator model proposed in \cite{zapletal2022stabilization}. Notably, for this method the mapping from the evolution of the density matrix to stochastic differential equations is without approximations, allowing quantum observables to be calculated exactly in cases that are numerically stable. By replacing the full density matrix evolution with an ensemble of stochastic trajectories, we circumvent the prohibitive memory requirements of the master equation. We demonstrate this capability with system sizes of $N = 21$ sites, although the method can be scaled to $N > 100$ easily. We focus specifically on the ''linear case" architecture, where the nonlinearity is provided by engineered local two-photon loss rather than Kerr interactions. This setup is of particular interest due to its predicted fast convergence to the steady-state manifold and its robustness against single-photon loss. 

The system consists of dissipatively coupled linear resonators subject to a two-photon drive, featuring both single-photon and two-photon loss (shown in Fig.~\ref{fig:fig_0}). It is fully described by the Lindblad master equation:

\begin{equation}\label{Eq:i1}
\begin{aligned}
    \dot{\hat{\rho}} = -i[\hat{H},\hat{\rho}] +(&\kappa_1 \sum_{j=1}^N \mathcal{D}[\hat{a}_j] +\kappa_2 \sum_{j=1}^N \mathcal{D}[\hat{a}^2_j] \\&+ \gamma \sum_{j=1}^N \mathcal{D}[\hat{a}_j - e^{i\phi}\hat{a}_{j+1}])\hat{\rho},
\end{aligned}
\end{equation}
where $\kappa_1$ and $\kappa_2$ are the amplitudes of single-photon and two-photon loss, respectively, $\gamma$ is the amplitude of non-local dissipation, and dissipator $\mathcal{D}[\hat{O}]=\hat{O}\hat{\rho}\hat{O}^{\dagger}-\tfrac{1}{2}\hat{O}^{\dagger}\hat{O}\hat{\rho}-\tfrac{1}{2}\hat{\rho}\hat{O}^{\dagger}\hat{O}$. $\hat{H}$ is the two-photon drive Hamiltonian given by:

\begin{equation}\label{Eq:i2}
    \hat{H} = \sum_{j=1}^N [\epsilon_j \hat{a}^{\dagger}_j\hat{a}^{\dagger}_j + \epsilon_j^*\hat{a}_j\hat{a}_j ],
\end{equation}
where $\epsilon_j = \epsilon e^{-i\theta j}$ is the amplitude of the two-photon drive, with a phase determined by the site $j$, and $\theta = 2\phi$.

Beyond enabling simulations of large resonator arrays, this approach reveals nonclassical correlations between opposite-quasimomentum modes. As highlighted in Fig.~\ref{fig:m7}, we find violations of the classical Cauchy--Schwarz bound in dissipatively coupled arrays containing $21$ and $41$ resonators. The definition of the correlation measure and a detailed analysis of these violations are presented in Sec. \ref{sec:Quasimomentum space}. Our simulations successfully reconstruct the dynamical behavior of local and global observables for large $N$, corroborating the scalability of the dissipation engineering protocol. We observe, however, that while population dynamics and lower-order moments are efficiently captured, the estimation of the parity is prone to simulation instability. We discuss the convergence properties of the stochastic method in different parameter regimes for both single-mode and multimode systems. Our results provide a first numerical glimpse into the dynamics of massively multimode cat states, bridging the gap between analytical predictions for thermodynamic limits and small-scale exact simulations.

\section{Results}

\subsection{Positive-$P$ representation of cat state systems}

The master equation in Eq.~(\ref{Eq:i1}) can be transformed into a Fokker-Planck equation governing the evolution of the distribution $P$ for the stochastic variables $\vec{\alpha}$ and $\vec{\beta}$, corresponding roughly to coherent state amplitudes of a ket and bra state, respectively. Samples of these variables allow for the estimation of observables and the reconstruction of the density matrix in the limit $s\rightarrow\infty$, where $s$ denotes the number of samples. The resulting stochastic differential equations for these trajectories are given by (see Sec.~\ref{ppderiv})
\begin{subequations}
\begin{equation}
\begin{aligned}
    \frac{d\alpha_j}{dt} &= (- \kappa_2 \alpha^2_j -2i\epsilon_j)\beta_j - (\gamma+\frac{\kappa_1}{2})\alpha_j + \frac{\gamma}{2}e^{i\phi}\alpha_{j+1} \\&+ \frac{\gamma}{2}e^{-i\phi}\alpha_{j-1} + \sqrt{- \kappa_2\alpha^2_j -2i\epsilon_j}\xi_j(t),
\end{aligned}
\end{equation}
\begin{equation}
\begin{aligned}
    \frac{d\beta_j}{dt} &= (- \kappa_2 \beta^2_j +2i\epsilon^*_j)\alpha_j - (\gamma+ \frac{\kappa_1}{2})\beta_j + \frac{\gamma}{2}e^{-i\phi}\beta_{j+1} \\&+ \frac{\gamma}{2}e^{i\phi}\beta_{j-1} + \sqrt{-\kappa_2\beta^2_j+2i\epsilon^*_j}\tilde{\xi}_j(t).
\end{aligned}
\end{equation}
\end{subequations}
The real-valued stochastic processes $\xi_j(t)$ and $\tilde{\xi}_j(t)$ are independent white-noise terms with zero mean, satisfying the correlations $\langle \xi_j(t) \xi_k(t')\rangle_s = \delta_{jk}\delta(t-t')$, $\langle \tilde{\xi}_j(t) \tilde{\xi}_k(t')\rangle_s = \delta_{jk}\delta(t-t')$, and $\langle \xi_j(t) \tilde{\xi}_k(t')\rangle_s = 0$. The terms $\xi_j(t)dt$ and $\tilde{\xi}_j(t)dt$ correspond to standard Wiener process increments and are numerically realized as Gaussian random variables with variance $1/\Delta t$, where $\Delta t$ is the simulation time step.

Normally ordered observables can be estimated from the stochastic samples using the relation
\begin{equation}
    \langle (\hat{a}^{\dagger})^m \hat{a}^n \rangle = \lim_{s\rightarrow\infty} \langle \beta^m \alpha^n\rangle_s.
\end{equation}
The full density matrix of the state can be reconstructed from the stochastic samples via
\begin{equation}\label{rhoelements}
    \hat{\rho} = \lim_{s\rightarrow\infty}\langle\hat{\Lambda}(\vec{\alpha},\vec{\beta})\rangle_s,
\end{equation}
where $\hat{\Lambda}(\vec{\alpha},\vec{\beta})$ is the kernel operator defined in Eq.~(\ref{Eq_1.31}). However, the primary advantage of the positive-$P$ method lies in its ability to efficiently estimate observables, whereas we observe that reconstructing the full density matrix remains computationally expensive in general.

\subsection{Transient dynamics of single-mode cat states}

In previous work, similar phase-space methods, such as the complex-$P$ representation, have been utilized to analytically derive the exact steady-state density matrix for the single-mode version of these systems \cite{minganti2016exact, bartolo2016exact}. However, since the distribution in the complex-$P$ representation is generally neither real nor positive, the corresponding stochastic differential equations do not exist \cite{deuar2005first}. Consequently, direct simulation of the system dynamics using the complex-$P$ representation is precluded. Although single-mode cat-state systems can be efficiently simulated using the master equation, benchmarking the positive-$P$ method in this context provides valuable insights into its performance in multimode cat-state systems.

\begin{figure}[h!]
    \centering
    \includegraphics[width=\linewidth]{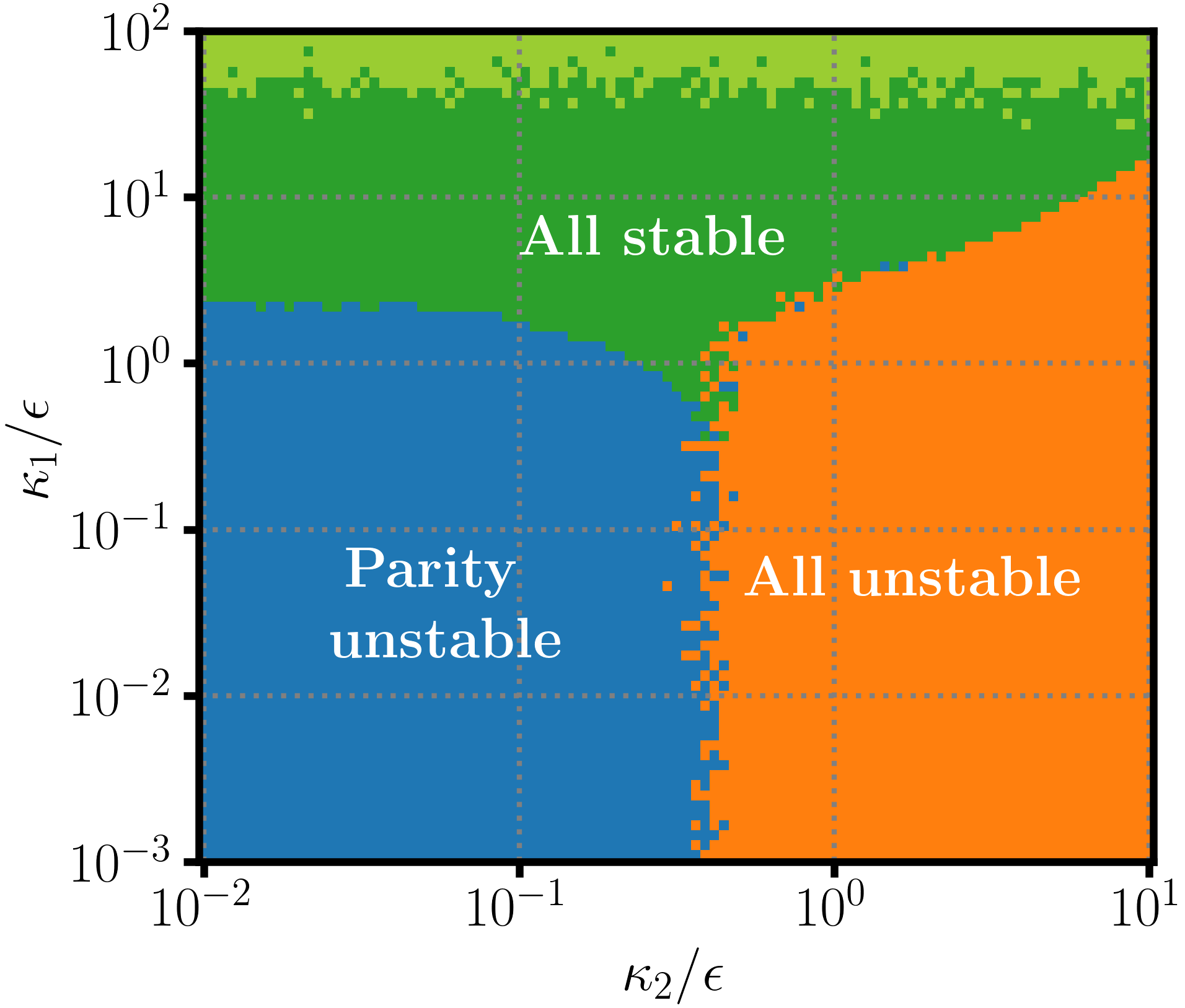}
    \caption{Stability plot of positive-\(P\) simulations for single-mode systems: orange region, unstable \(\alpha\) and \(\beta\) trajectories before the persistent state; blue region, unexpected parity decay with all other observables well behaved; green region, stable and reliable behavior of all observables; yellow-green region, low signal-to-noise ratio of the second-order correlation function due to an indeterminate zero-by-zero form as the average photon number approaches zero in the large single-photon-dissipation limit.}
    \label{fig:positive_p_phase}
\end{figure}

To evaluate the method's performance in single-mode systems, we set $N = 1$ and $\gamma=0$ in Eq.~(\ref{Eq:i1}). Previous studies have shown that with localized interactions and losses, useful positive-P simulation times are typically determined by the behavior of single modes or small local subsystems \cite{deuar2006first,deuar2006pp,deuar2021fully}. In the single-mode case, the relevant observables include the average photon number $n$, the coherent amplitude $\zeta$, the second-order correlation function $g_2$, and the parity $\Pi$. Their expressions in terms of the operators $\hat{a}$ and $\hat{a}^{\dagger}$, as well as the stochastic variables $\alpha$ and $\beta$, are given by

\begin{subequations}\label{Eq:linear_observables_pp}
\begin{equation}
        n = \langle \hat{a}^{\dagger} \hat{a} \rangle = \lim_{s\rightarrow\infty} \text{Re}\{\langle\alpha\beta\rangle_s\}
\end{equation}
\begin{equation}
    \zeta = \sqrt{\langle \hat{a}^2\rangle}=\lim_{s\rightarrow \infty}\sqrt{\langle\alpha^2\rangle_s}
\end{equation}
\begin{equation}
    g_2 = \frac{\langle \hat{a}^{\dagger} \hat{a}^{\dagger} \hat{a} \hat{a} \rangle}{\langle \hat{a}^{\dagger} \hat{a}\rangle ^2} = \lim_{s\rightarrow\infty}\frac{\text{Re}\{\langle \alpha^2\beta^2\rangle_s\}}{n^2}
\end{equation}
\begin{equation}
    \Pi =\langle \exp{(i \pi \hat{a}^{\dagger} \hat{a}})\rangle = \lim_{s\rightarrow\infty}\text{Re}\{\langle\exp{(-2\alpha\beta)}\rangle_s\}
\end{equation}
\end{subequations}

\begin{figure*}
    \centering
    \includegraphics[width=\linewidth]{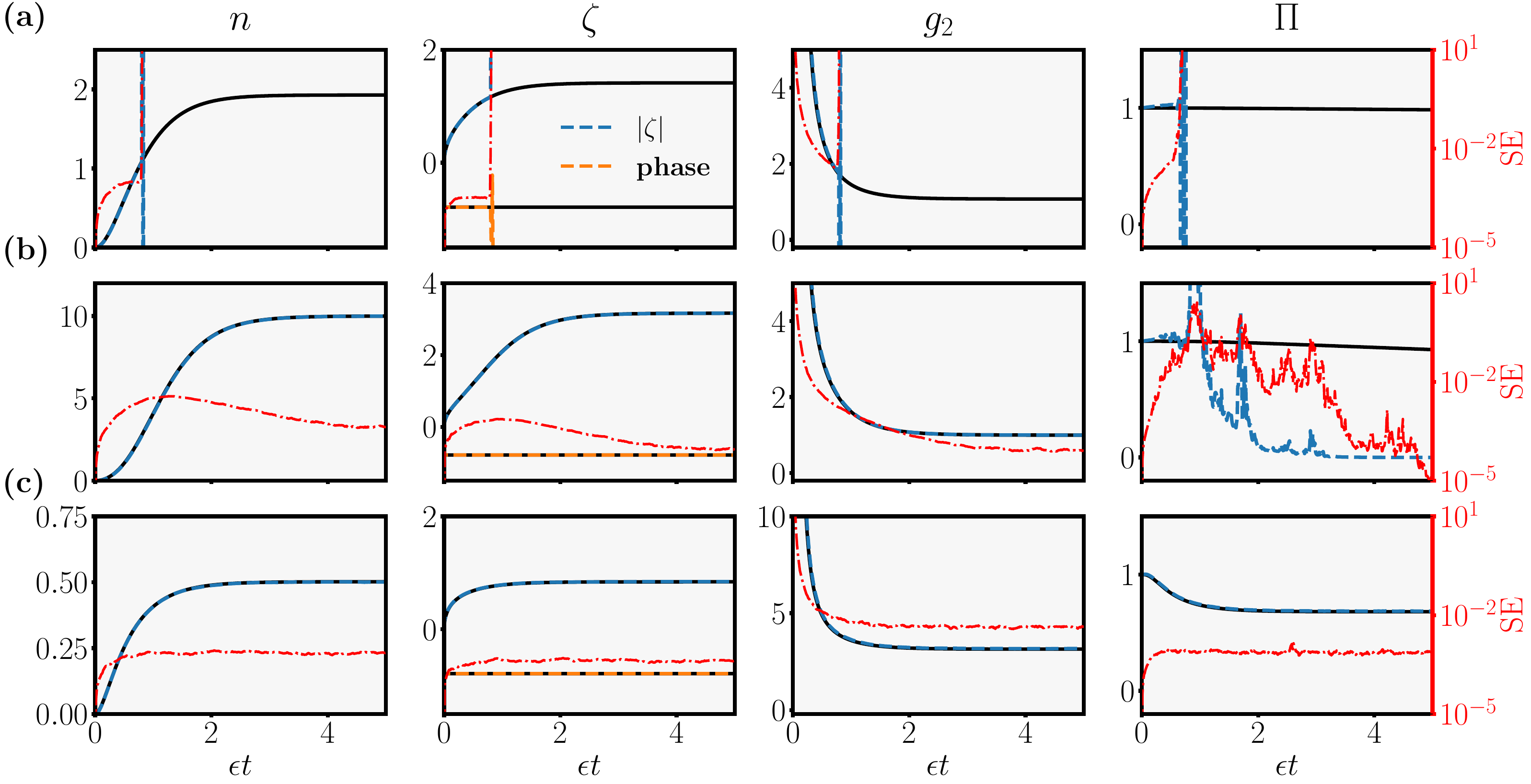}
    \caption{Comparison of single-mode observable dynamics from the master equation (black dashed lines) and the positive-\(P\) method (solid colored lines), with expectation values in blue, the phase of \(\zeta\) in orange, and statistical errors (SE) in red on a logarithmic scale. (a) \(\kappa_1/\epsilon=10^{-3}\), \(\kappa_2/\epsilon=1\): unstable stochastic trajectories before the persistent state. (b) \(\kappa_1/\epsilon=10^{-3}\), \(\kappa_2/\epsilon=0.2\): unexpected parity decay, with excellent agreement for all other observables. (c) \(\kappa_1/\epsilon=5\), \(\kappa_2/\epsilon=0.2\): excellent agreement for all observables.
    }
    \label{fig:positive_p_phase_details}
\end{figure*}

\begin{figure}
    \centering
    \includegraphics[width=\linewidth]{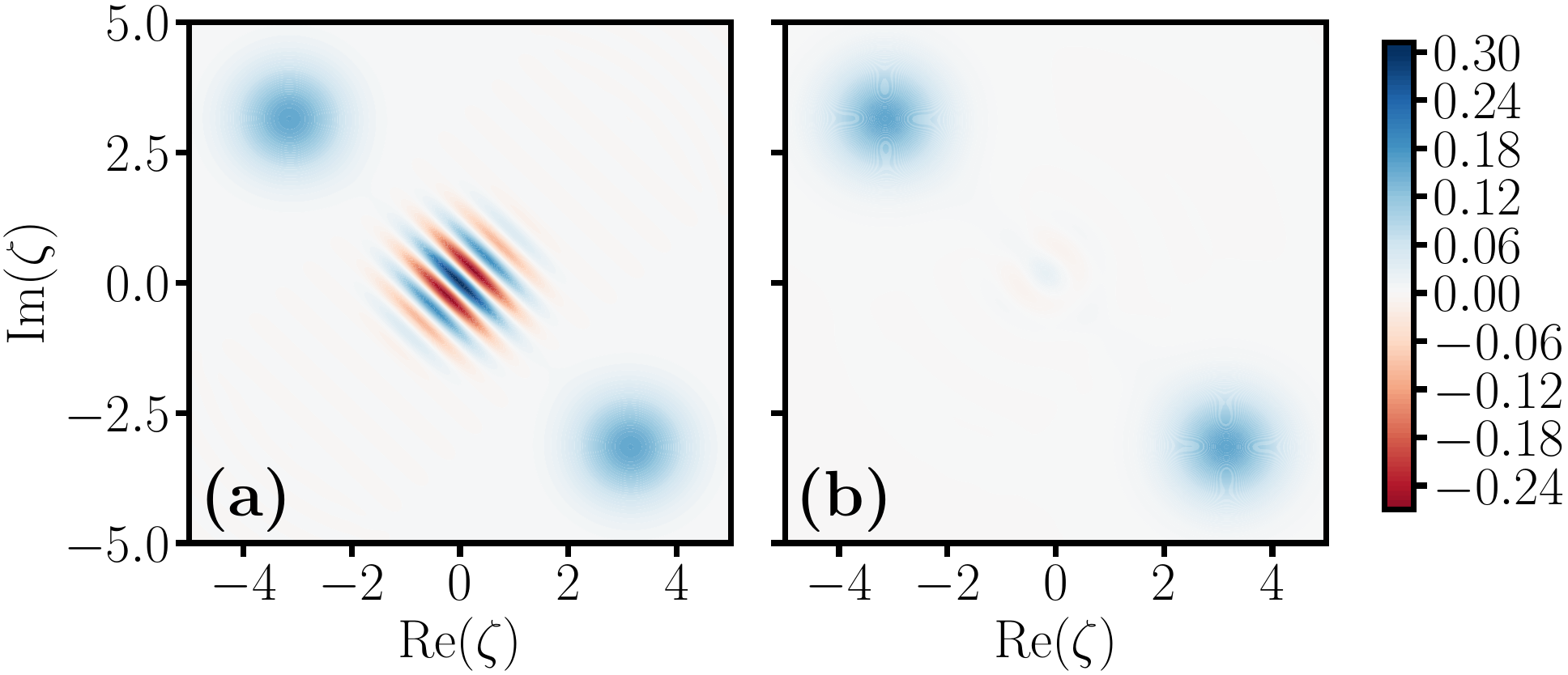}
    \caption{Wigner functions of the single-mode density matrix from (a) the master equation and (b) the positive-\(P\) simulation. Parameters: \(\kappa_1/\epsilon=10^{-3}\), \(\kappa_2/\epsilon=0.2\), and \(\epsilon t=3\), corresponding to transient cat-state generation. Ensemble size: \(10^6\) trajectories.}
    \label{fig:wigner_positive_p}
\end{figure}

We identify three distinct parameter regimes that determine the stability of the simulation when initialized in the vacuum state and evolving toward a persistent cat-like  state (hereafter referred to as ``persistent state"), which is not the true steady state but an exceptionally long-lived transient. In the orange region of Fig.~\hyperref[fig:positive_p_phase]{\ref*{fig:positive_p_phase}}, where the two-photon loss $\kappa_2$ is large, the stochastic trajectories become unstable before any persistent state is reached (see Fig.~\hyperref[fig:positive_p_phase_details]{\ref*{fig:positive_p_phase_details}(a)}). This unstable regime encompasses a subset of the parameter space that would otherwise yield clearly resolved cat states.

Clearly resolved cat states can also be generated with smaller $\kappa_2$ (blue region). In this regime, we find that all observables, with the exception of parity, agree with the master equation results (see Fig.~\hyperref[fig:positive_p_phase_details]{\ref*{fig:positive_p_phase_details}(b)}). The parity undergoes an unexpected decay, suggesting that information regarding the state's parity is lost before the persistent state is obtained. Notably, both the parity value and its statistical error exhibit spikes (known indicators of numerical issues \cite{gilchrist1997positive,deuar2002gauge}), and their magnitudes become comparable as the simulation begins to lose stability. The parity of a cat state corresponds to the value of its Wigner function at the origin, $W(0,0)$. In Fig.~\hyperref[fig:wigner_positive_p]{\ref*{fig:wigner_positive_p}}, we compare the Wigner function obtained from the master equation with the state reconstructed from the positive-$P$ simulation for a representative example in this regime. We observe that in the positive-$P$ simulation, the central interference fringes, which indicate the state's parity, are absent. Consequently, we identify that the density matrix recovered from the positive-$P$ simulation represents a classical mixture of two coherent states, taking the form of
\begin{equation}
\begin{aligned}
    \rho &= \frac{1}{2}(|C_{\zeta}^+\rangle\langle C_{\zeta}^+| + |C_{\zeta}^-\rangle\langle C_{\zeta}^-|)\\
    &= \frac{1}{2}(|\zeta\rangle\langle\zeta| + |-\zeta\rangle\langle-\zeta|).
\end{aligned}
\end{equation}
By contrast, in the regime characterized by a large single-photon loss rate (green area), all observables exhibit good agreement with the master equation results (corresponding to Fig.~\hyperref[fig:positive_p_phase_details]{\ref*{fig:positive_p_phase_details}(c)}). However, this parameter regime is of limited interest, as it yields only highly damped or vacuum states in single-mode reductions of this model.

The extreme sensitivity of the parity observable, compared to standard observables such as photon number or field quadratures, stems directly from its mathematical expression. Standard observables are computed from low-order polynomials of the stochastic variables. In contrast, the parity operator requires evaluating an exponential function $\exp(i\pi \hat{a}^{\dagger}\hat{a})$. Large deviations experienced by the stochastic variables are exponentially amplified when the parity observable is evaluated. Consequently, the parity observable develops a heavy-tailed distribution, where its value is determined by very few escaping trajectories. This causes the occurrence of boundary errors of the second kind, leading to the observed spikes and decay (Supplementary Note 1). Addressing the underlying limitation, the failure to capture this high-order quantum interference (the central Wigner fringes) represents a known breakdown of the standard \mbox{positive-$P$} representation for highly non-classical states. While fundamental to the standard method, these diverging trajectories can potentially be mitigated by alternative sampling strategies, such as stochastic gauge methods (e.g., the gauge-$P$ representation \cite{deuar2002gauge}), which we explore for the single-mode case in Supplementary Note 2, albeit at the cost of increased computational complexity. We find that appropriate drift-gauge functions can be constructed to modify the deterministic part of the stochastic differential equations, thereby improving the stability of the single-mode simulations.

\subsection{Observables and stability in multimode systems}
In multimode systems, in addition to the local observables defined in Eq.~\ref{Eq:linear_observables_pp}, we examine several non-local observables: the first-order $g^{(1)}(j,j')$ and second-order $g^{(2)}(j,j')$ spatial correlation functions between sites $j$ and $j'$, and the global parity $\Pi$ of the multimode state. Their definitions are given by

\begin{subequations}\label{Eq:linear_observables_pp_multimode}
\begin{equation}
g_1(j,j')=\frac{\langle\hat{a}_{j}^{\dagger}\hat{a}_{j'}\rangle}{\sqrt{\langle\hat{a}_{j}^{\dagger}\hat{a}_{j}\rangle\langle\hat{a}_{j'}^{\dagger}\hat{a}_{j'}\rangle}} = \lim_{s\to\infty}\frac{\langle\alpha_{j'}\beta_{j}\rangle_s}{\sqrt{n(j)n(j')}}
\end{equation}
\begin{equation}
g_2(j,j')=\frac{\langle\hat{a}_{j}^{\dagger}\hat{a}_{j'}^{\dagger}\hat{a}_{j}\hat{a}_{j'}\rangle}{\langle\hat{a}_{j}^{\dagger}\hat{a}_{j}\rangle\langle\hat{a}_{j'}^{\dagger}\hat{a}_{j'}\rangle}=\lim_{s\to\infty}\frac{\langle\alpha_j\alpha_{j'}\beta_j\beta_{j'}\rangle_s}{n(j)n(j')}
\end{equation}
\begin{equation}
\Pi = \langle \exp(i\pi\sum_j\hat{a}_j^{\dagger}\hat{a}_j)\rangle = \lim_{s\to\infty}\langle\exp(-2\sum_j\alpha_j\beta_j)\rangle_s.
\end{equation}
\end{subequations}

We assess the effectiveness of the positive-$P$ method in multimode systems by comparing it with quantum trajectory simulations for a small system with $N=3$ sites (see Supplementary Note 3). With the exception of the global parity, the local and non-local observables show excellent agreement between the two methods. In particular, although the local parity is problematic in the corresponding single-mode system, it is accurately captured in the multimode case near the Zeno limit. This stability arises from the large amplitude of the non-local dissipator $\gamma$, which plays a role similar to that of $\kappa_1$ in stabilising the stochastic trajectories. However, the global multimode parity is the only observable for which the two methods exhibit a significant discrepancy. Similar to the single-mode case, it displays an unexpected decay before the persistent state is reached. Since the global parity estimator contains stochastic variables from all modes in the exponential, it is extremely sensitive to rare fluctuations and boundary-term errors. The accumulation of these errors becomes more severe as the number of modes increases, while lower-order local and non-local observables remain well behaved. We also applied the stochastic drift gauges developed for the single-mode case, but observed no significant improvement in simulation stability due to the additional coupling terms in the multimode stochastic equations (see Supplementary Note 3).

\begin{figure}[t]
    \centering
    \includegraphics[width=1.0\linewidth]{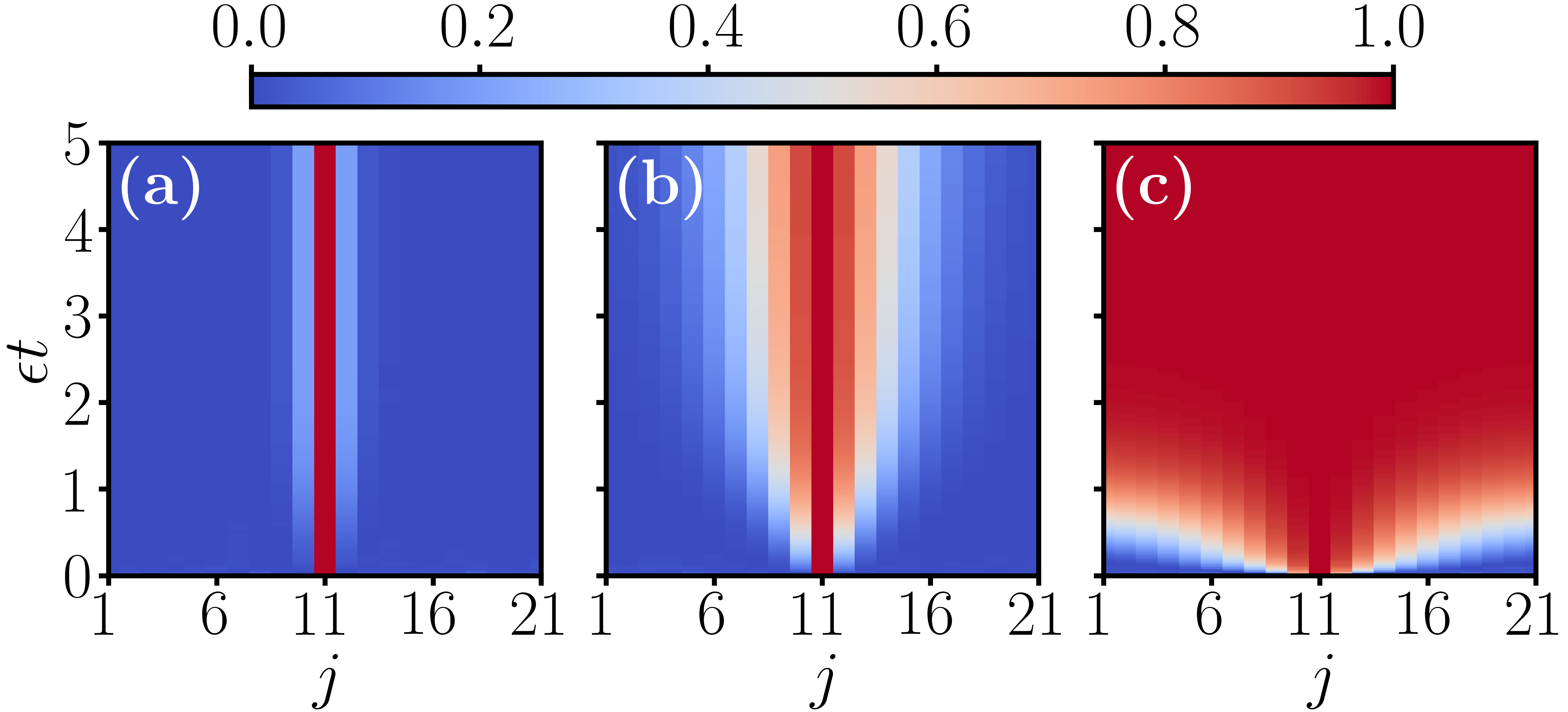}
    \caption{Time evolution of the first-order correlation function \(\lvert g_1(11,j)\rvert\) between reference site 11 and site \(j\) for \(N=21\) modes. Columns: (a) \(\gamma/\epsilon=0.2\), (b) \(\gamma/\epsilon=2\), and (c) \(\gamma/\epsilon=50\), with increasing non-local dissipation.}
    \label{fig:m2}
\end{figure}

\subsection{Spatial correlations}

Simulating the dynamics of multimode systems using the master equation or quantum trajectories can be computationally prohibitive. The positive-$P$ method significantly reduces both memory and time complexity, enabling the efficient prediction of observables for large-scale systems across a broad range of parameters~\cite{deuar2021fully}. We leverage this capability to investigate spatial correlations under varying strengths of non-local dissipation. Hereafter, we adopt parameters that reflect standard operating regimes in modern superconducting-circuit implementations: $\kappa_1/\epsilon=0.001$ and $\kappa_2/\epsilon=0.2$. In such engineered platforms, single-photon loss must be strongly suppressed relative to the driven two-photon processes to maintain the robust noise bias of the cat states, a condition routinely achieved in experiments (see, e.g., Refs.~\cite{grimm2020stabilization,leghtas2015confining}). For these parameters, we set the phase difference to $\phi=0$ and simulate a system with $N=21$ spatial modes. The first- and second-order spatial correlation functions provide direct insight into the coherence and statistical properties of the system. The first-order spatial correlation function, $g^{(1)}(j,j')$, quantifies the coherence between sites $j$ and $j'$. It is a normalised quantity with a magnitude ranging from zero to one. The second-order spatial correlation function, $g^{(2)}(j,j')$, characterises the correlations between particles detected at sites $j$ and $j'$.

Fig.~\ref{fig:m2} shows the time evolution of $\lvert g^{(1)}(11,j)\rvert$ between the reference site $11$ and all other sites for different strengths of the non-local dissipation. When weak non-local dissipation is introduced, for example $\gamma/\epsilon=0.2$, with a magnitude comparable to the two-photon loss rate $\kappa_2$, a slow and limited propagation of phase coherence is observed (Fig.~\hyperref[fig:m2]{\ref*{fig:m2}(a)}). The first-order correlation function develops a peak around the reference site and decays with distance, indicating a finite but short coherence length. As the non-local dissipation is increased to $\gamma/\epsilon=2$, a clear ballistic light cone emerges in the propagation of $g^{(1)}$ (Fig.~\hyperref[fig:m2]{\ref*{fig:m2}(b)}). Although full coherence is not yet established across the entire system, the correlations propagate over a much longer distance, indicating the development of long-range order. In the Zeno limit, where the non-local dissipation dominates, for example $\gamma/\epsilon=50\gg\kappa_1/\epsilon,\kappa_2/\epsilon$, the propagation velocity of the first-order coherence increases significantly (Fig.~\hyperref[fig:m2]{\ref*{fig:m2}(c)}). The light cone rapidly encompasses the entire array, and the persistent state develops $g^{(1)}\approx1$ across all spatial modes. This indicates the formation of a globally coherent state resembling a Bose--Einstein condensate. In this regime, the generation of high-fidelity multimode cat states is predicted in Ref.~\cite{zapletal2022stabilization}.

The second-order spatial correlations show additional bunching and antibunching features as the non-local dissipation is increased. For weak non-local dissipation, the correlations between distinct sites remain negligible. At intermediate coupling, strong on-site bunching and short-range antibunching emerge, likely due to the competition between the coherent drive dynamics and dissipative coupling. In the Zeno limit, the second-order correlation function becomes nearly uniform and remains slightly above unity. The corresponding dynamical results and spatial distributions are presented in Supplementary Note 4.

\subsection{Quasimomentum space}
\label{sec:Quasimomentum space}

In the quasimomentum space defined within the first Brillouin zone, a plane-wave basis is given by
\begin{equation}\label{Eq: fourier}
    \hat{b}_k = (1/\sqrt{N})\sum_{j=1}^{N}\exp(ijk)\hat{a}_j,
\end{equation}
where the quasimomentum $k$ takes values ranging from $2\pi/N$ to $2\pi$ in steps of $2\pi/N$. In the Zeno limit, the system can be described by an effective master equation derived using first-order perturbation theory in momentum space \cite{zapletal2022stabilization}:

\begin{equation}
    \dot{\hat{\rho}}_{\phi} = \frac{\kappa_2}{N}\mathcal{D}[\hat{b}^2_{\phi}-\tilde{\zeta}^2]\hat{\rho}_{\phi}.
\end{equation}
In this regime, only a single ``dark mode" with momentum $k=\phi$ is populated and remains dissipation-free, while all other modes experience strong dissipation and are rapidly projected onto vacuum states. A limitation of this approximation is that it describes only systems with very large $\gamma$. For smaller values of $\gamma$, multiple momentum modes may become occupied. Consequently, capturing the properties of the full system requires including the Liouvillian terms describing these other modes. Here, we map the momentum domain from $(0,2\pi]$, as defined in previous work \cite{zapletal2022stabilization}, to the symmetric interval $(-\pi,\pi]$ by re-centering $k$ as
\begin{equation}
    k_{\text{new}} =
    \begin{cases}
    k - 2\pi, & \text{if } k > \pi \\
    k, & \text{otherwise.}
    \end{cases}
\end{equation}

\begin{figure}[h!]
    \centering
    \includegraphics[width=\linewidth]{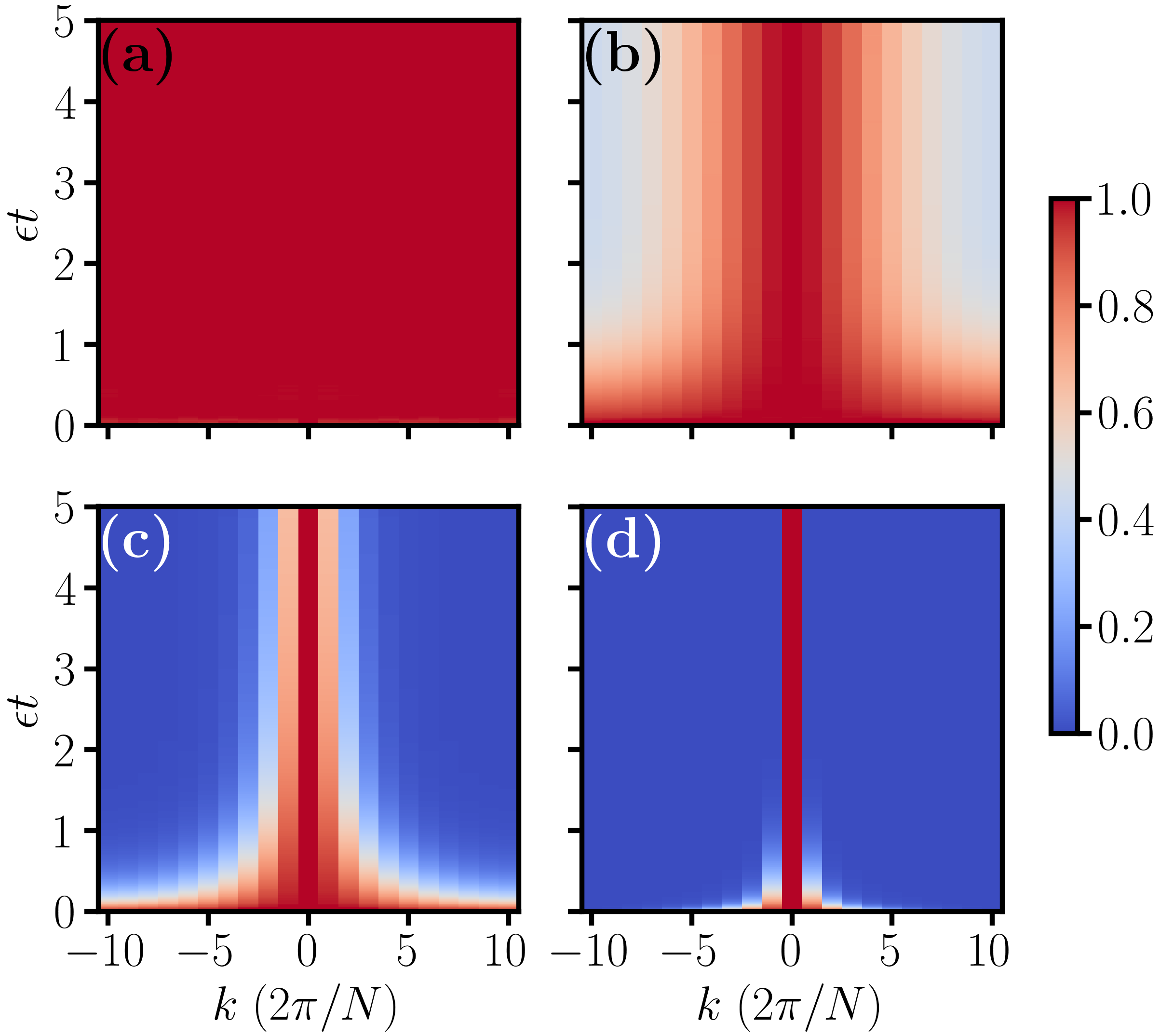}
    \caption{Time evolution of the normalized average photon number $\langle \hat{n}_k\rangle/\langle\hat{n}_{\phi}\rangle$ in momentum space. The subplots correspond to (a) $\gamma/\epsilon=0$, (b) $\gamma/\epsilon=0.2$, (c) $\gamma/\epsilon=2$, and (d) $\gamma/\epsilon=50$.}
    \label{fig:m4}
\end{figure}

For each momentum mode $k$, we utilize the stochastic variables derived from Eq.~(\ref{Eq: fourier}) to estimate observables in momentum space. In Fig.~\hyperref[fig:m4]{\ref*{fig:m4}}, we display the normalized average photon number $\langle \hat{n}_k\rangle/\langle\hat{n}_{\phi}\rangle$ for various modes in momentum space across different values of $\gamma$. From (a) to (d), the ratio $\gamma/\epsilon$ is gradually increased from 0 (decoupled) to 50 (the Zeno limit). The choice of $\phi$ determines the dark mode (the sole populated mode in the Zeno limit). Since we set $\phi = 0$, the ``zero-momentum" mode is selected as the dark mode and remains populated throughout. In (a), due to the absence of dissipative coupling, all momentum modes are evenly occupied. As non-local dissipation is introduced, we observe that the photon numbers in the $k\neq \phi$ modes eventually vanish. As $\gamma$ increases, the system progressively suppresses photon occupation in the less favored lossy modes. In the ideal Zeno limit shown in (d), only the $k=0$ mode remains populated.

Directly quantifying entanglement in the system using the positive-$P$ method is challenging, as it cannot be extracted from a single observable measurement (one might try with tomography of the density matrix via (\ref{rhoelements}) but this tends to be noisy). However, the antipropagating second-order correlation functions in momentum space, $\tilde{g}_2(k,-k)$, serve as powerful indicators of entanglement between bosons. We investigate $\tilde{g}_2(k,-k)$ across the transition from the decoupled regime to the Zeno limit.

\begin{figure}[t]
    \centering
    \includegraphics[width=\linewidth]{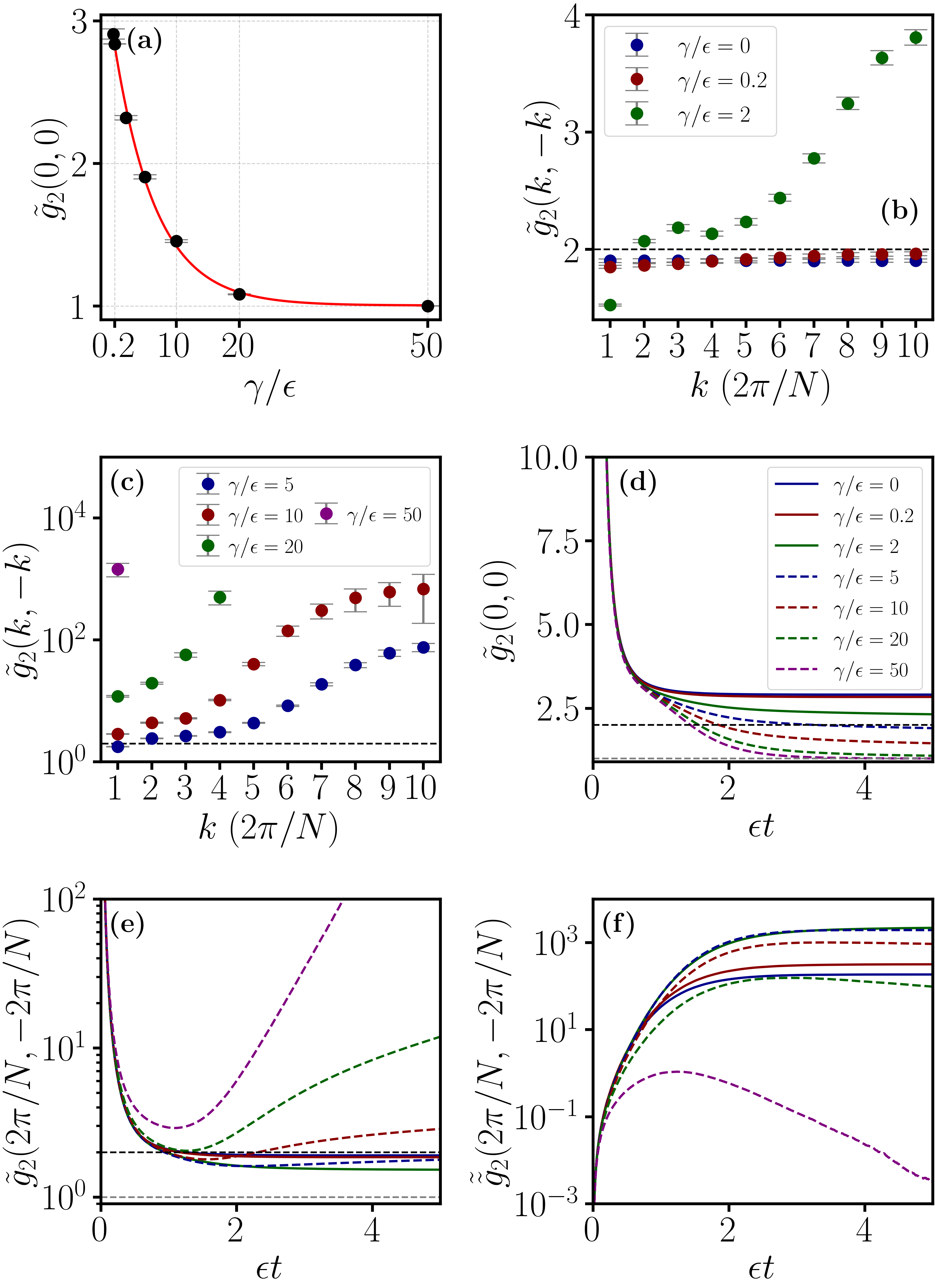}
    \caption{Momentum-space second-order correlations \(\tilde{g}_2\) between opposite-sign modes \(k\) and \(-k\) in the \(N=21\)-mode system for various values of \(\gamma\). (a) Dark mode (\(k=\phi\)) at \(\epsilon t=5\). (b, c) Equal-amplitude dissipative modes of opposite sign. Omission of data for \(k>4\) at \(\gamma/\epsilon=20\) and \(k>1\) at \(\gamma/\epsilon=50\) due to low signal-to-noise ratios. (d, e) Time evolution of \(\tilde{g}_2\) for the dark mode and the mode closest to the dark mode, respectively. (f) Time evolution of the unnormalized second-order correlation function \(\tilde{\tilde{g}}_2\).}
    \label{fig:m5}
\end{figure}

First, we focus on the dark mode. In Fig.~\hyperref[fig:m5]{\ref*{fig:m5}(a)}, we find that the value of $\tilde{g}_2(0,0)$ of the persistent states at $\epsilon t = 5$ decreases exponentially with $\gamma$, following the relation
\begin{equation}
    \tilde{g}_2(0,0) = 1.9\;\text{exp}(-0.15 \gamma/\epsilon) + 1.
\end{equation}
Figure \hyperref[fig:m5]{\ref*{fig:m5}(d)} shows the time evolution of $\tilde{g}_2(0,0)$, confirming that it reaches a steady value at $\epsilon t = 5$ for all values of $\gamma$. In the absence of non-local dissipation, the system develops strong photon bunching in the dark mode ($\tilde{g}_2(0,0) \sim 3$). Conversely, in the Zeno limit, we expect the formation of a multimode cat state, which in momentum space has a wavefunction of the form \cite{zapletal2022stabilization} (with $\phi=0$):
\begin{equation}
    |\Psi\rangle = \mathcal{N}_{\pm}(|\zeta\rangle_{\phi}\pm|-\zeta\rangle_{\phi}) \otimes (\bigotimes_{k\neq\phi}|0\rangle_k).
\end{equation}
In this case, we find that $\tilde{g}_2(0,0)$ approaches unity, indicating that the photons follow Poissonian statistics in the dark mode. This characteristic of a coherent-state-like distribution provides evidence for the formation of a high-photon-number single-mode cat state predicted in \cite{grimm2020stabilization} within the dark mode. For the $k\neq \phi$ momentum modes at $\epsilon t =5$, we observe that when $\gamma$ is small, all $\tilde{g}_2(k,-k)$ values remain slightly below 2 ($\sim 1.9$) (Fig.~\hyperref[fig:m5]{\ref*{fig:m5}(b)}), indicating weaker bunching than that of a perfectly thermal state. This suggests that while the system is dominated by chaotic behavior, a small degree of coherence persists and competes with these chaotic properties. For large $\gamma$, all non-zero momentum $\tilde{g}_2$ values rise above 2, indicating super-bunching behavior (Fig.~\hyperref[fig:m5]{\ref*{fig:m5}(c)}). This implies that the bosons are more strongly correlated than in a perfect thermal state. Such behavior is a strong signature of a non-equilibrium persistent state containing particle pairs, which may indicate the presence of strong quantum correlations and entanglement.

In Fig.~\hyperref[fig:m5]{\ref*{fig:m5}(e)}, we find that the $\tilde{g}_2$ of the non-zero momentum modes does not approach a steady value but grows continuously with time after reaching a minimum. This behavior arises from a vanishing denominator, where the normalization ($n_{\pm k}$) in the denominator of $\tilde{g}_2$ decreases much faster than the unnormalized second-order correlation function $\tilde{\tilde{g}}_2$. In the Zeno limit ($\gamma/\epsilon=50$), we confirmed that the non-zero momentum modes decay rapidly to zero, as shown in Fig.~\hyperref[fig:m4]{\ref*{fig:m4}(d)}. In Figs.~\hyperref[fig:m5]{\ref*{fig:m5}(e,f)}, we observe that while the unnormalized $\tilde{\tilde{g}}_2$ decreases exponentially to $\sim 10^{-3}$, the normalized $\tilde{g}_2$ grows exponentially. These are behaviors characteristic of vacuum modes occupied almost purely by a small density of entangled pairs.

In this weakly populated regime, large values of the normalized
correlation $\tilde{g}_2$ can arise from the small mode populations
and do not by themselves quantify entanglement. To test the
nonclassicality of the intensity correlations between distinct
opposite-quasimomentum modes, we use the normally ordered
Cauchy--Schwarz parameter \cite{kheruntsyan2012violation},
\begin{equation}
    R_{\mathrm{CS}}(k,-k)
    =
    \frac{
        \left|\left\langle:\hat{n}_k\hat{n}_{-k}:\right\rangle\right|
    }{
        \sqrt{
            \left\langle:\hat{n}_k^2:\right\rangle
            \left\langle:\hat{n}_{-k}^2:\right\rangle
        }
    }.
\end{equation}
Here, colons denote normal ordering, with
$:\hat{n}_q^2:
=\hat{b}_q^{\dagger 2}\hat{b}_q^2
=\hat{n}_q(\hat{n}_q-1)$.
For distinct modes $k$ and $-k$, the cross-correlation is already
normally ordered, so that
$\langle:\hat{n}_k\hat{n}_{-k}:\rangle
=\langle\hat{n}_k\hat{n}_{-k}\rangle$.
Classical fields admitting a nonnegative Glauber--Sudarshan
$P$ representation satisfy
\begin{equation}
    \left|\left\langle:\hat{n}_k\hat{n}_{-k}:\right\rangle\right|
    \le
    \sqrt{
        \left\langle:\hat{n}_k^2:\right\rangle
        \left\langle:\hat{n}_{-k}^2:\right\rangle
    }.
\end{equation}
A statistically significant value of $R_{\mathrm{CS}}(k,-k)>1$
therefore witnesses nonclassical intensity correlations.

For $\gamma=0$, the Cauchy--Schwarz parameter remains equal to unity for all opposite-momentum mode pairs (not shown). As highlighted in Fig.~\ref{fig:m7}, finite dissipative coupling produces violations of the classical bound, with $R_{\mathrm{CS}}(k,-k)>1$ for nonzero opposite-quasimomentum mode pairs. Here we examine these correlations in detail. This violation of the Cauchy--Schwarz inequality demonstrates a departure from classical behavior, indicating the presence of nonlocal quantum correlations. Notably, the violation is far stronger than seen in the referenced BEC collision experiments \cite{kheruntsyan2012violation} and the signal remains well above noise and therefore presumably measurable even in the higher $k$ values studied.

\begin{figure}
    \centering
    \includegraphics[width=\linewidth]{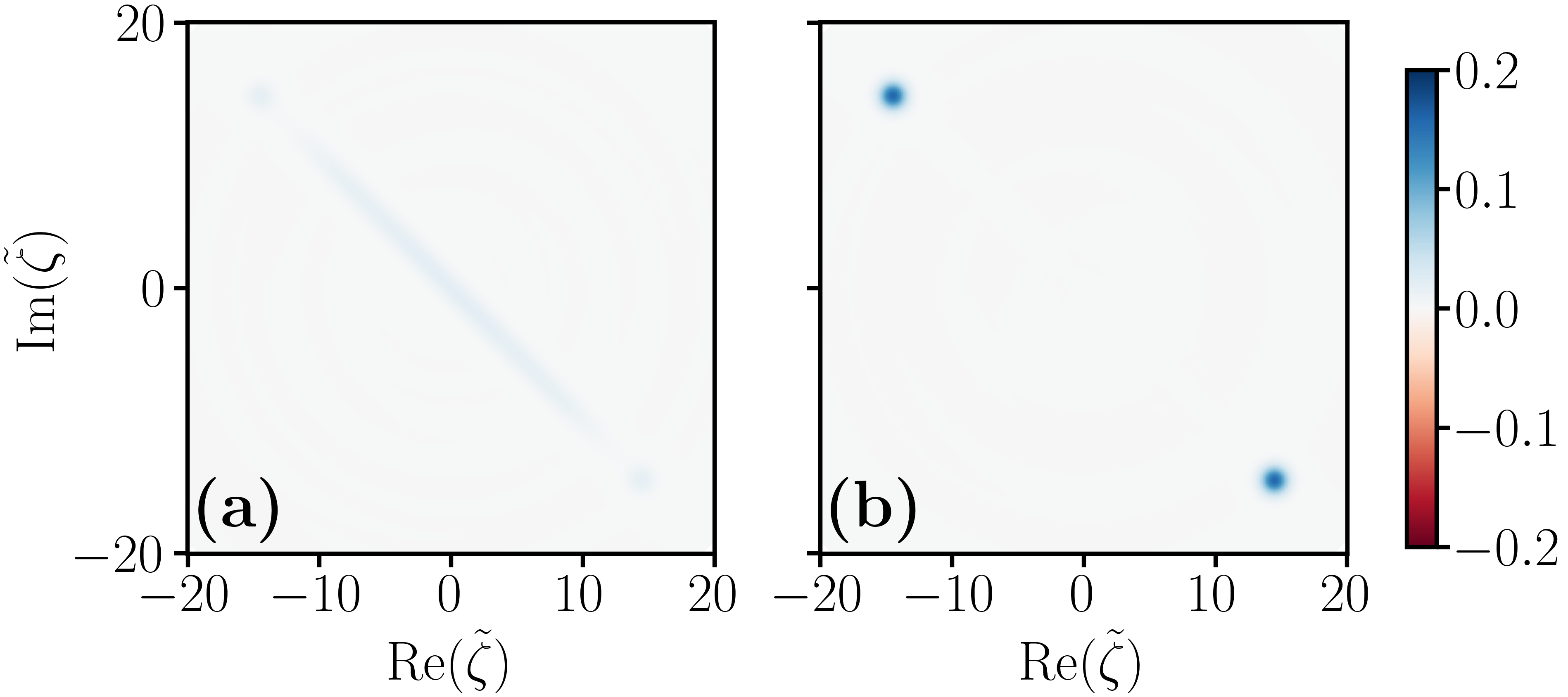}
    \caption{Wigner function plots of the dark mode state for (a) $\gamma/\epsilon = 2$ and (b) $\gamma/\epsilon = 50$.}
    \label{fig:m6}
\end{figure}

We further characterize the dark mode states by reconstructing their density matrices. In Fig.~\ref{fig:m6}, we plot the Wigner functions of the dark mode states obtained from the positive-$P$ simulations. For the small-coupling regime ($\gamma/\epsilon=2$), the Wigner function exhibits a single central interference fringe, indicating positive parity. For the Zeno-limit case ($\gamma/\epsilon=50$), we observe patterns consistent with those from single-mode positive-$P$ simulations: the Wigner function displays two distinct local maxima without interference fringes, suggesting that the parity of the multimode state cannot be estimated reliably in this regime.

\section{Discussion}

In this work, we successfully implemented the positive-$P$ method to simulate the stabilization of multimode cat states, effectively overcoming the memory bottleneck associated with conventional approaches such as master equation or quantum trajectory simulations. We achieved stable and convergent simulations for realistic system parameters in arrays up to $N = 21$ sites, a system size that far surpasses the previous limit of $N = 3$ accessible via the master equation simulation. Moreover, the computational cost of our method scales linearly with system size, making simulations of $N > 100$ sites feasible. Our approach yields accurate estimates for a wide range of observables, including photon numbers, coherent amplitudes, and both spatial and quasi-momentum space correlation functions. These capabilities confirm the positive-$P$ method as a valid and powerful tool for characterizing dissipative couplings in large-scale quantum systems.

However, we identify a specific limitation regarding the estimation of the state parity. We find that the parity observable is prone to divergent sampling error in the positive-$P$ representation, affecting both single-mode and multimode simulations. While stochastic gauge techniques can stabilize these trajectories in single-mode cases, their generalization to the multimode linear case presents significant challenges. The complexity of the multimode drift terms implies that standard gauge choices may fail to reduce, or even increase, the sampling noise. Despite this, the non-classical nature and coherence of the stabilized states are partially captured through the computed correlation functions and further supported by the observed violation of the Cauchy--Schwarz inequality. Furthermore, we explored the system across various parameter regimes, ranging from weak to strong dissipative coupling. Unlike previous analytical works that rely on an effective single-mode master equation valid only in the Zeno limit, our numerical method captures the full dynamics across all coupling strengths, providing a complete picture of the transition into the protected manifold.

\section{Methods}

\subsection{Positive-$P$ representation of the system}
\label{ppderiv}

Here, we present the detailed derivation of the positive-$P$ representation dynamics for this system, following a route explained in  \cite{deuar2021fully,drummond1980generalised,deuar2005first}. The density matrix $\hat{\rho}$ of the bosonic modes can be expressed as
\begin{equation}\label{Eq_1.29}
    \hat{\rho} = \int d^{2N}\! \vec{\alpha}\, d^{2N}\! \vec{\beta}\, P(\vec{\alpha},\vec{\beta})\hat{\Lambda}(\vec{\alpha},\vec{\beta}),
\end{equation}
where $P(\vec{\alpha},\vec{\beta})$ is the positive-$P$ distribution of the complex variables $\vec{\alpha}$ and $\vec{\beta}$, and $\hat{\Lambda}(\vec{\alpha},\vec{\beta})$ is the kernel operator. For each mode, this operator is defined by
\begin{equation}\label{Eq_1.31}
    \hat{\Lambda}_j(\alpha_j,\beta_j) = \frac{|\alpha_j\rangle\langle \beta^*_j|}{\langle \beta^*_j|\alpha_j\rangle}.
\end{equation}
The states $|\alpha_j\rangle$ and $|\beta_j\rangle$ are the Bargmann coherent states \cite{deuar2006first}, given by
\begin{equation}
    |\alpha\rangle = e^{-\frac{1}{2}|\alpha|^2} e^{\alpha \hat{a}^{\dagger}}|0\rangle.
\end{equation}
We can rewrite Eq.~(\ref{Eq_1.31}) as
\begin{equation}\label{Eq_1.32}
    \hat{\Lambda}_j(\alpha_j,\beta_j) = e^{-\alpha_j\beta_j}e^{\alpha_j\hat{a}^{\dagger}_j}|0_j\rangle\langle 0_j|e^{\beta_j\hat{a}_j}.
\end{equation}
Since coherent states are left/right eigenstates of the annihilation ($\hat{a}$) and creation ($\hat{a}^{\dagger}$) operators, respectively, the following relations hold:
\begin{subequations}
    \begin{equation}\label{Eq_1.33}
    \hat{a}_j\hat{\Lambda}_j = \alpha_j\hat{\Lambda}_j,
\end{equation}
\begin{equation}\label{Eq_1.34}
    \hat{\Lambda}_j\hat{a}^{\dagger}_j = \beta_j\hat{\Lambda}_j.
\end{equation}
\end{subequations}
Taking the derivatives with respect to $\alpha_j$ and $\beta_j$ of Eq.~(\ref{Eq_1.32}), we obtain:
\begin{subequations}
    \begin{equation}
    \hat{\Lambda}_j\hat{a}_j = (\alpha_j+\frac{\partial}{\partial \beta_j})\hat{\Lambda}_j,
\end{equation}
\begin{equation}
    \hat{a}^{\dagger}_j\hat{\Lambda}_j = (\beta_j+\frac{\partial}{\partial \alpha_j})\hat{\Lambda}_j.
\end{equation}
\end{subequations}
These identities allow us to map the master equation to the Fokker-Planck equation. For a given Hamiltonian $\hat{H}$ and dissipator $\mathcal{D}[\hat{\mathcal{O}}]$ with strength $\gamma$, the system is described by the Lindblad master equation
\begin{equation}\label{Eq_master_equation}
    \dot{\hat{\rho}} = -i[\hat{H},\hat{\rho}]+\gamma\mathcal{D}[\hat{\mathcal{O}}]\hat{\rho},
\end{equation}
where $\mathcal{D}[\hat{\mathcal{O}}]\hat{\rho} = \hat{\mathcal{O}}\hat{\rho}\hat{\mathcal{O}}^{\dagger}-\frac{1}{2}\hat{\mathcal{O}}^{\dagger}\hat{\mathcal{O}}\hat{\rho}-\frac{1}{2}\hat{\rho}\hat{\mathcal{O}}^{\dagger}\hat{\mathcal{O}}$. Substituting Eq.~(\ref{Eq_1.29}) into the master equation yields:
\begin{equation}
    \begin{aligned}
    &\int d^{4N}\!\vec{v}\,\frac{\partial P}{\partial t}\hat{\Lambda} = \\ &\int d^{4N}\!\vec{v}\,P[\sum_{u}A_u\frac{\partial}{\partial u}+\sum_{u,v}\frac{1}{2}D_{u,v}\frac{\partial^2}{\partial u\partial v}]\hat{\Lambda},
    \end{aligned}
\end{equation}
where $u,v\in\{\vec{\alpha},\vec{\beta}\}$. Assuming the positive-$P$ distribution decays sufficiently fast as $\alpha,\beta \rightarrow \infty$, we can neglect the boundary terms to obtain:
\begin{equation}
    \begin{aligned}
           &\int d^{4N}\!\vec{v}\,\frac{\partial P}{\partial t}\hat{\Lambda} = \\ &\int d^{4N}\!\vec{v}\,\hat{\Lambda}[-\sum_u\frac{\partial}{\partial u}A_u+\sum_{u,v}\frac{1}{2}\frac{\partial^2}{\partial u\partial v}D_{u,v}]P. 
    \end{aligned}
\end{equation}
Since the trace of the kernel operator is $\text{Tr}[\hat{\Lambda}] = 1$, we derive the Fokker-Planck equation:
\begin{equation}
    \frac{\partial P}{\partial t} = -\sum_u\frac{\partial}{\partial u}(A_u P)+\sum_{u,v}\frac{1}{2}\frac{\partial^2}{\partial u \partial v}(D_{u,v}P).
\end{equation}
The Fokker-Planck equation for our model is given by
\begin{equation}
    \begin{aligned}
        \frac{\partial{P}}{\partial{t}} &= 
        - \sum_j \frac{\partial}{\partial{\alpha_j}}[(- \kappa_2 \alpha^2_j  -2i\epsilon_j)\beta_j\\
        &+ \frac{\gamma}{2}e^{i\phi}\alpha_{j+1} + \frac{\gamma}{2}e^{-i\phi}\alpha_{j-1} - ({\gamma}+\frac{\kappa_1}{2})\alpha_j]P\\
        &- \sum_j \frac{\partial}{\partial{\beta_j}}[(- \kappa_2  \beta^2_j +2i\epsilon_j^*)\alpha_j\\
        &+ \frac{\gamma}{2}e^{-i\phi}\beta_{j+1} + \frac{\gamma}{2}e^{i\phi}\beta_{j-1} - (\gamma+\frac{\kappa_1}{2})\beta_j]P\\
        &+\sum_j \frac{1}{2}\frac{\partial^2}{\partial \alpha^2_j}[-\kappa_2\alpha^2_j-2i\epsilon_j]P\\
        &+\sum_j \frac{1}{2}\frac{\partial^2}{\partial \beta^2_j}[- \kappa_2\beta^2_j+2i\epsilon_j^*]P,
    \end{aligned}
\end{equation}
This yields the $j$-th elements of the drift vector $\mathbf{A}$ and diffusion matrix $\mathbf{D}$:
\begin{equation}
    A_j = \begin{pmatrix}
\begin{aligned}
(-\kappa_2\alpha_j^2-2i\epsilon_j)\beta_j &-\Big(\gamma+\tfrac{\kappa_1}{2}\Big)\alpha_j \\
  + \tfrac{\gamma}{2} e^{i\phi}\alpha_{j+1}
  &+ \tfrac{\gamma}{2} e^{-i\phi}\alpha_{j-1} \\
  &
\end{aligned} \\
\begin{aligned}
(-\kappa_2\beta_j^2+2i\epsilon_j^*)\alpha_j &-\Big(\gamma+\tfrac{\kappa_1}{2}\Big)\beta_j \\
  + \tfrac{\gamma}{2} e^{-i\phi}\beta_{j+1}
  &+ \tfrac{\gamma}{2} e^{i\phi}\beta_{j-1} 
\end{aligned}
\end{pmatrix}
\end{equation}
\begin{equation}\label{Eq. D1}
    D_{j} = \begin{pmatrix} -\kappa_2\alpha^2_j-2i\epsilon_j & 0 \\ 0 & - \kappa_2\beta^2_j+2i\epsilon_j^* \end{pmatrix}.
\end{equation}
The corresponding stochastic differential equation is
\begin{equation}
    \frac{\partial \vec{v}}{\partial t} = A(\vec{v}) + B(\vec{v}) \vec{\xi}(t),
\end{equation}
where $\vec{\xi}(t)$ represents uncorrelated real Gaussian white noise satisfying $\langle \xi_u(t)\xi_v(t')\rangle_s = \delta(t-t')\delta_{u,v}$, and $B$ is the matrix square root of $D$ satisfying $D = BB^{T}$. Since $D$ is diagonal, a natural choice for $B$ is
\begin{equation}\label{Eq. D2}
    B_{j} = \begin{pmatrix} \sqrt{-\kappa_2\alpha^2_j-2i\epsilon_j} & 0 \\ 0 & \sqrt{- \kappa_2\beta^2_j+2i\epsilon_j^*} \end{pmatrix}.
\end{equation}

\subsection{Comparison between positive-$P$ and master equation results}
To obtain the results presented in Fig.~\hyperref[fig:positive_p_phase]{\ref*{fig:positive_p_phase}}, we fixed the driving amplitude at $\epsilon = \epsilon^* = 1$ and varied $\kappa_1$ and $\kappa_2$ to calculate the steady-state solution of the master equation using the steady-state solver in QuTiP \cite{johansson2012qutip,JOHANSSON20131234}. In parallel, we simulated the stochastic variables $\alpha$ and $\beta$ starting from a vacuum initial condition. The simulations ran until either (i) the observables obtained from the positive-$P$ simulation matched their steady-state values, or (ii) the simulation became unstable before any agreement was reached. The different parameter regimes were then identified by comparing the final values of the positive-$P$ simulation with the master equation results.

In certain cases, we observed an unexpected decay of the parity observable to zero during the simulation (blue on Fig.~\ref{fig:positive_p_phase} with example in Fig.~\hyperref[fig:positive_p_phase_details]{\ref*{fig:positive_p_phase_details}(b)}). A simple comparison between the final value of the positive-$P$ simulation and the steady-state solution is insufficient to detect this discrepancy, as the steady state of the cat-state system is typically a mixture of two opposite-parity cat states, it also possesses zero parity. 
We confirm that a reliable method to identify this discrepancy when one does not have access to the exact time-dependent results for comparison is to monitor for the presence of spikes in the observable, indicating the occurrence of boundary term errors as per \cite{gilchrist1997positive}.

In the large single-photon loss limit, the second-order correlation function encounters a $0/0$ situation as both the numerator and denominator vanish. By measuring the signal-to-noise ratio, we identify the regime (SNR = $g_2/\sigma_{SE} < 3$, yellow-green) in Fig.~\ref{fig:positive_p_phase} where large statistical errors make correlation calculations impractical. Notably, increasing the number of stochastic trajectories for select parameter points in this regime yields no significant improvement in the SNR.

\subsection{Estimation of statistical errors}

Regarding the parameter sweep in Fig.~\ref{fig:positive_p_phase}, the statistical error (SE) is obtained by sampling $10^5$ trajectory pairs per parameter set, partitioned into $s = 20$ subensembles. In all other simulations, we sample $10^6$ pairs and divide them into $s = 100$ subensembles. For each subensemble $j$, the observable value $O_j$ is calculated according to Eq.~(\ref{Eq:linear_observables_pp}). The final estimate is then given by $\langle \hat{O}\rangle = O \pm \sigma_{SE}$, where the mean is defined as $O = \sum_j O_j /s$ and the standard error as $\sigma_{SE} = \sqrt{\text{var}(O_j)/(s-1)}$.

\subsection{Writing Assistance}

During the preparation of this work, the authors used Gemini (Google) to assist with drafting and refining the Introduction section, as well as improving the writing of other parts of the manuscript to enhance readability and structure. After using this tool, the authors reviewed and edited the content as needed and take full responsibility for the content of the publication.


\section*{Data Availability}
The data generated and analyzed during the current study are available from the corresponding author upon reasonable request.

\vspace{1em}

\section*{Code Availability}
The code used to generate the data in this study is available from the corresponding author upon reasonable request.

\section*{Acknowledgements}
This work was supported by the Engineering and Physical Sciences Research Council [grant number EP/S021582/1]. E.G. acknowleges support from the Engineering and Physical Sciences Research Council [grant number EP/Z000505/1]. M.S. and A.F. acknowlege support from the Engineering and Physical Sciences Research Council [grant number EP/V026496/1] and from the EUCENTRAL project, funded by the European Union under Horizon Europe (project No. 101186579).  Y.S., E.G., M.S. acknowledge support from the QuantERA project PROTEQT.

\section*{Competing Interests}
The authors declare no competing interests.

\section*{Author Contributions}
The derivation, numerical simulations, and manuscript preparation were carried out by Y.S., under the supervision and guidance of A.F., P.D., E.G., and M.S. P.D. also contributed to revising and improving the manuscript. All authors discussed the results and the manuscript.


\bibliography{citations}

\end{document}


\title{Supplementary Materials: Stochastic phase-space simulation of multimode cat states via positive-$P$ representation}

\author{Yi Shi}
\affiliation{Department of Physics and Astronomy, University College London, Gower Street, London WC1E 6BT, United Kingdom}

\author{Alex Ferrier}
\affiliation{Department of Physics and Astronomy, University College London, Gower Street, London WC1E 6BT, United Kingdom}
\affiliation{Center for Theoretical Physics, Polish Academy of Sciences, Aleja Lotnik\'ow 32/46, 02-668 Warsaw, Poland}

\author{Piotr Deuar}
\affiliation{Institute of Physics, Polish Academy of Sciences,
Aleja Lotników 32/46, PL-02-668 Warsaw, Poland}

\author{Eran Ginossar}
\affiliation{School of Mathematics and Physics,
University of Surrey, Guildford GU2 7XH, United Kingdom}

\author{Marzena Szymanska}
\affiliation{Department of Physics and Astronomy, University College London, Gower Street, London WC1E 6BT,
United Kingdom}

\maketitle

\section{Supplementary Note 1. Stability analysis of the single-mode case}

\begin{figure}
    \centering
    \includegraphics[width=\linewidth]{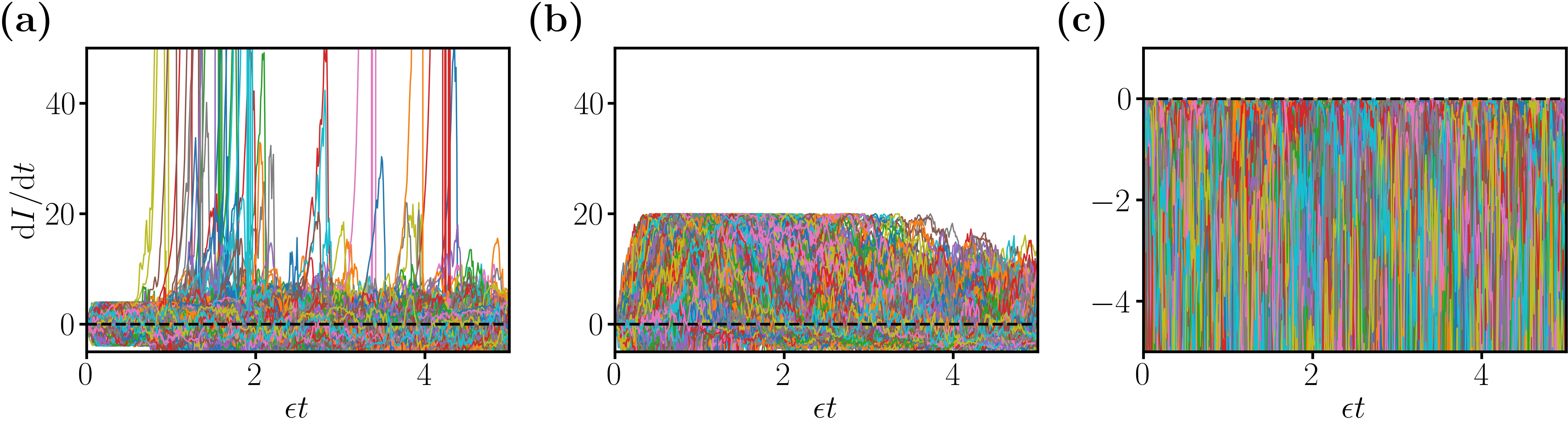}
    \caption{Time derivative $\text{d}I/\text{d}t$ for 1000 pairs of stochastic trajectories. Parameter sets: (a) $\kappa_1/\epsilon = 10^{-3}, \kappa_2/\epsilon = 1$; (b) $\kappa_1/\epsilon = 10^{-3}, \kappa_2/\epsilon = 0.2$; and (c) $\kappa_1/\epsilon = 5, \kappa_2/\epsilon = 0.2$.}
    \label{fig:dIdt}
\end{figure}

In this section, we discuss the mechanisms governing the partial failure of stochastic simulations. Let us define a function $I$ as the sum of the squared magnitudes of $\alpha$ and $\beta$:
\begin{equation}
    I(\alpha,\beta) = |\alpha|^2 + |\beta|^2.
\end{equation}
The function $I$ describes the distance between the trajectory and the origin in the complex phase space. Considering only the drift component of the stochastic equations, the time derivative of $I$ is given by
\begin{equation}
    \text{d}I/\text{d}t = - 8 \epsilon \text{Im}(\alpha\beta^*) - \kappa_1 I - 2\kappa_2 I \text{Re}(\alpha\beta).
\end{equation}
Numerical instability typically arises when $\text{d}I/\text{d}t > 0$. In our case, three scenarios emerge:

\begin{itemize}
  \item \textbf{Scenario 1, Global Simulation Failure:} As shown in Fig. 4(a) of the main text, instability in all observables occurs before any transient state is reached. This instability arises from boundary term errors of the first kind, as discussed in \cite{deuar2005first}, where the trajectories become highly unstable. When $\text{d}I/\text{d}t > 0$, the stochastic trajectories tend to diverge rapidly. In Fig. \hyperref[fig:dIdt]{\ref*{fig:dIdt}(a)}, the time derivative $\text{d}I/\text{d}t$ is not bounded by zero and grows exponentially during the early stages of the simulation. Consequently, the exact mapping from the Lindblad master equation to the Fokker-Planck equation breaks down, as the probability of finding these trajectories at infinity is non-zero.
\end{itemize}

\begin{itemize}
  \item \textbf{Scenario 2, Selective Observable Failure:} In Fig. 4(b) of the main text, only the parity observable fails to converge. Here, the system is not strictly bounded by $\text{d}I/\text{d}t \le 0$ (Fig. \hyperref[fig:dIdt]{\ref*{fig:dIdt}(b)}), indicating that boundary-term errors of the first kind still occur. However, they are less severe than in Scenario 1, as the rate of divergence is bounded. Nevertheless, the distribution $P$ can develop pathological, slowly decaying tails. Since the parity observable scales exponentially with $\alpha$ and $\beta$, boundary-term errors of the second kind \cite{deuar2005first} may arise. Intuitively, while most sampled values are clustered near zero, the state's parity is dominated by rare outlier events located far from the origin (Fig. \hyperref[fig:S34]{\ref*{fig:S34}(a)}), resulting in a poor signal-to-noise ratio. Although the exponential dependence of the parity observable amplifies these numerical errors, other observables remain insensitive and can be estimated accurately.

\end{itemize} 

\begin{itemize}
  \item \textbf{Scenario 3, Stable Simulations:} In Fig. 4(c), the parity and other observables show convergence and excellent agreement with the master equation. In this regime, the system is strictly bounded by $\text{d}I/\text{d}t \le 0$ (Fig. \hyperref[fig:dIdt]{\ref*{fig:dIdt}(c)}), preventing boundary-term errors. This stability arises because the $-\frac{\kappa_1}{2}\alpha(\beta)$ drift term, which acts as a restoring force, has a sufficiently large amplitude.
\end{itemize}

Boundary-term errors can be mitigated by introducing stochastic gauges. In the following section, we explore two types of stochastic drift gauges. Although these gauges eliminate boundary errors in Scenario 1 and yield excellent agreement for the parity observable, they compromise simulation stability in Scenario 2.

\section*{Supplementary Note 2. Stochastic drift gauges}\label{Supplementary Note 2. Stochastic drift gauges}

To resolve boundary-term errors, stochastic drift gauges can be introduced to replace repulsive terms with restoring forces \cite{deuar2002gauge}. This is achieved by identifying the terms causing divergence and replacing them with terms dependent on the magnitude of the stochastic variables, effectively driving the system toward stability. Arbitrary gauge functions $g_k$ are added to the Langevin equations:

\begin{equation}\label{Eq:linear_drift_gauges}
    d x_j = A_j dt + \sum_k B_{jk}(dW_k-g_k dt)
\end{equation}
where $x_j \in \{\alpha,\beta \}$.

Stochastic gauges are added to the drift component of the stochastic differential equations \cite{deuar2005first}, dynamically shifting the center of the random step for each trajectory. Consequently, a weight function $\Omega$ must be introduced to compensate for these drift modifications. The weight function evolves according to
\begin{equation}
    d\Omega = \Omega[A_0 dt + \sum_k g_k dW_k],
\end{equation}
where $A_0$ is the Stratonovich correction (zero in the Itô calculus). The corrected density matrix and observables are obtained via:

\begin{equation}
    \begin{aligned}
        \hat{\rho} &= \lim_{s\rightarrow\infty} \langle\Omega \Lambda(\alpha,\beta)\rangle_s\\
        \langle(\hat{a}^{\dagger})^m (\hat{a})^n\rangle &= \lim_{s\rightarrow\infty} \langle\Omega \alpha^n\beta^m\rangle_s\\.
    \end{aligned}
\end{equation}
Thus, the observables are modified as follows:

\begin{equation}\label{Eq:linear_observables_gp}
    \begin{aligned}
        n &= \langle \hat{a}^{\dagger} \hat{a} \rangle = \lim_{s\rightarrow\infty} \langle\Omega\alpha\beta\rangle_s\\
        \zeta &= \sqrt{\langle \hat{a}^2\rangle}=\lim_{s\rightarrow \infty}\sqrt{\langle\Omega\alpha^2\rangle_s}\\
        g_2 &= \frac{\langle \hat{a}^{\dagger} \hat{a}^{\dagger} \hat{a} \hat{a} \rangle}{\langle \hat{a}^{\dagger} \hat{a}\rangle ^2} = \lim_{s\rightarrow\infty}\frac{\langle \Omega\alpha^2\beta^2\rangle_s}{(\langle\Omega\alpha\beta\rangle_s)^2}\\
        \Pi &=\langle \exp{(i \pi \hat{a}^{\dagger} \hat{a}})\rangle = \lim_{s\rightarrow\infty}\langle\Omega\{ \exp{(-2\alpha\beta)}\}\rangle_s.
    \end{aligned}
\end{equation}
\subsection{2.1 Two different drift gauges}
For the single-mode system, the simplified Fokker-Planck equation is given by

\begin{equation}
    \begin{aligned}
        \frac{\partial{P}}{\partial{t}} =&\\
        &- \frac{\partial}{\partial{\alpha}}(-2i \beta \epsilon -\kappa_1 \alpha/2 -\kappa_2\alpha^2\beta)P - \frac{\partial}{\partial{\beta}}(2i\alpha\epsilon^*-\kappa_1\beta/2-\kappa_2\beta^2\alpha)P\\
        &+\frac{1}{2}\frac{\partial^2}{\partial \alpha^2}(-2i\epsilon - \alpha^2\kappa_2)P+\frac{1}{2}\frac{\partial^2}{\partial \beta^2}(2i\epsilon^*-\beta^2\kappa_2)P.
    \end{aligned}
\end{equation}
The drift vector $A$ is formulated as
\begin{equation}\label{Eq:linear_pp_A}
    A = \begin{pmatrix}
        -2i \beta \epsilon -\kappa_1 \alpha/2 -\kappa_2\alpha^2\beta\\
        2i\alpha\epsilon^*-\kappa_1\beta/2-\kappa_2\beta^2\alpha,
        \end{pmatrix}
\end{equation}
while the diffusion matrix $D$ is expressible as
\begin{equation}\label{Eq:linear_pp_D}
    D = \begin{pmatrix}
        -2i\epsilon - \alpha^2\kappa_2 & 0 \\
        0 & 2i\epsilon^*-\beta^2\kappa_2
        \end{pmatrix}.
\end{equation}
The noise matrix $B$ in the corresponding stochastic differential equation (SDE) represents a decomposition of the diffusion matrix, such that $D = BB^{T}$. The stochastic simulations in the main text are based on the SDE using $B = \sqrt{D}$, given by
\begin{equation}\label{Eq:linear_pp_B}
    B = \begin{pmatrix}
        \sqrt{-2i\epsilon - \alpha^2\kappa_2} & 0 \\
        0 & \sqrt{2i\epsilon^*-\beta^2\kappa_2}
        \end{pmatrix}.
\end{equation}

We seek a gauge function that applies a restoring force to diverging trajectories. We define a gauge function similar to Eq. (61) in \cite{deuar2002gauge}, adding a term of the form $-f(x_j) x_j$ to the drift. Here, $f(x_j)$ must be quadratic and positive to neutralise the cubic terms in Eq. (\ref{Eq:linear_pp_A}). Implementing the modifications from \cite{deuar2002gauge} directly appears challenging due to the high non-linearity of the matrix elements in Eq. (\ref{Eq:linear_pp_B}). However, we can decompose the matrix $D$ in Eq. (\ref{Eq:linear_pp_D}) differently by isolating terms in each element. This yields a simpler noise matrix $B$, at the cost of introducing two additional independent noise terms:

\begin{equation}\label{Eq:linear_pp_B2}
    B = \begin{pmatrix}
        \sqrt{-2i\epsilon} & 0 & i\alpha\sqrt{\kappa_2} & 0 \\
        0 & \sqrt{2i\epsilon} & 0 & i\beta\sqrt{\kappa_2}
        \end{pmatrix}.
\end{equation}
This allows us to select the gauge functions
\begin{equation}\label{Eq:linear_pp_G2}
    G = \begin{pmatrix}
        \beta\sqrt{-2i\epsilon}\\
        \alpha\sqrt{2i\epsilon}\\
        i\sqrt{\kappa_2}(\alpha\beta-|\alpha\beta|)\\
        i\sqrt{\kappa_2}(\alpha\beta-|\alpha\beta|)
        \end{pmatrix},
\end{equation}
leading to a simplified drift vector $A$. In this form, the $\epsilon$ term cancels out, and the cubic term acts as a restoring force:
\begin{equation}\label{Eq:linear_gp_A2}
    A \rightarrow \begin{pmatrix}
        -\kappa_1\alpha/2-\kappa_2\alpha|\alpha\beta|\\
        -\kappa_1\beta/2-\kappa_2\beta|\alpha\beta|
        \end{pmatrix}.
\end{equation}

Although non-trivial, it is possible to use the noise matrix $B$ from Eq. (\ref{Eq:linear_pp_B}) directly and identify gauge functions that stabilise the stochastic trajectories. We define the set of gauge functions:

\begin{equation}\label{Eq:linear_pp_G}
    G = \begin{pmatrix}
        \beta\sqrt{-2i\epsilon-\kappa_2\alpha^2}+\alpha(\sqrt{-2i\epsilon-\kappa_2\alpha^2})^*\\
        \alpha\sqrt{2i\epsilon-\kappa_2\beta^2}+\beta(\sqrt{2i\epsilon-\kappa_2\beta^2})^*.
        \end{pmatrix}
\end{equation}
Combining this with Eqs. \ref{Eq:linear_pp_A}, \ref{Eq:linear_pp_B}, and \ref{Eq:linear_drift_gauges} yields:

\begin{equation}\label{Eq:linear_gp_A}
    A \rightarrow \begin{pmatrix}
        -\kappa_1\alpha/2-|2i\epsilon+\kappa_2\alpha^2|\alpha\\
        -\kappa_1\beta/2-|2i\epsilon-\kappa_2\beta^2|\beta
        \end{pmatrix}.
\end{equation}
In this instance, we introduce a stabilising term that includes $\epsilon$, rendering the term highly non-linear. Nevertheless, the absolute value ensures it functions as a restoring force. We summarise the two choices of the noise matrix $B$ and the corresponding drift gauges in Tab. \ref{tab:Comparison of Drift and Diffusion Matrices}.

\begin{table}[h!]
\makebox[\textwidth][c]{
\begin{tabular}{ccc}
\hline
\multicolumn{1}{|c|}{Choice 1}      & \multicolumn{1}{c|}{Positive-$P$}       & \multicolumn{1}{c|}{Gauge-$P$} \\ \hline
\multicolumn{1}{|c|}{Drift $A$}     & \multicolumn{1}{c|}{$\begin{pmatrix}
        -2i \beta \epsilon -\kappa_1 \alpha/2 -\kappa_2\alpha^2\beta\\
        2i\alpha\epsilon^*-\kappa_1\beta/2-\kappa_2\beta^2\alpha
        \end{pmatrix}$}               & \multicolumn{1}{c|}{$\begin{pmatrix}
        -\kappa_1\alpha/2-\kappa_2\alpha|\alpha\beta|\\
        -\kappa_1\beta/2-\kappa_2\beta|\alpha\beta|
        \end{pmatrix}$}      \\ \hline
\multicolumn{1}{|c|}{Diffusion $B$} & \multicolumn{2}{c|}{$\begin{pmatrix}
        \sqrt{-2i\epsilon} & 0 & i\alpha\sqrt{\kappa_2} & 0 \\
        0 & \sqrt{2i\epsilon} & 0 & i\beta\sqrt{\kappa_2}
        \end{pmatrix}$}                                              \\ \hline
\multicolumn{1}{|c|}{Gauge $G$}     & \multicolumn{1}{c|}{/}                & \multicolumn{1}{c|}{$\begin{pmatrix}
        \beta\sqrt{-2i\epsilon}\\
        \alpha\sqrt{2i\epsilon}\\
        i\sqrt{\kappa_2}(\alpha\beta-|\alpha\beta|)\\
        i\sqrt{\kappa_2}(\alpha\beta-|\alpha\beta|)
        \end{pmatrix}$}      \\ \hline
                                    &                                       &                              \\ \hline
\multicolumn{1}{|c|}{Choice 2}      & \multicolumn{1}{c|}{Positive-$P$}       & \multicolumn{1}{c|}{Gauge-$P$} \\ \hline
\multicolumn{1}{|c|}{Drift $A$}     & \multicolumn{1}{c|}{$\begin{pmatrix}
        -2i \beta \epsilon -\kappa_1 \alpha/2 -\kappa_2\alpha^2\beta\\
        2i\alpha\epsilon^*-\kappa_1\beta/2-\kappa_2\beta^2\alpha
        \end{pmatrix}$}               & \multicolumn{1}{c|}{$\begin{pmatrix}
        -\kappa_1\alpha/2-|2i\epsilon+\kappa_2\alpha^2|\alpha\\
        -\kappa_1\beta/2-|2i\epsilon-\kappa_2\beta^2|\beta
        \end{pmatrix}$}      \\ \hline
\multicolumn{1}{|c|}{Diffusion $B$} & \multicolumn{2}{c|}{$\begin{pmatrix}
        \sqrt{-2i\epsilon - \alpha^2\kappa_2} & 0 \\
        0 & \sqrt{2i\epsilon^*-\beta^2\kappa_2}.
        \end{pmatrix}$}                                              \\ \hline
\multicolumn{1}{|c|}{Gauge $G$}     & \multicolumn{1}{c|}{/} & \multicolumn{1}{c|}{$\begin{pmatrix}
        \beta\sqrt{-2i\epsilon-\kappa_2\alpha^2}+\alpha(\sqrt{-2i\epsilon-\kappa_2\alpha^2})^*\\
        \alpha\sqrt{2i\epsilon-\kappa_2\beta^2}+\beta(\sqrt{2i\epsilon-\kappa_2\beta^2})^*.
        \end{pmatrix}$}      \\ \hline
\end{tabular}}
\caption{Drift and diffusion matrices in positive-$P$ and gauge-$P$ representations.}
\label{tab:Comparison of Drift and Diffusion Matrices}
\end{table}

\subsection{2.2 Numerical results of transient dynamics}

\begin{figure}[h!]
    \centering
    \includegraphics[width=0.9\linewidth]{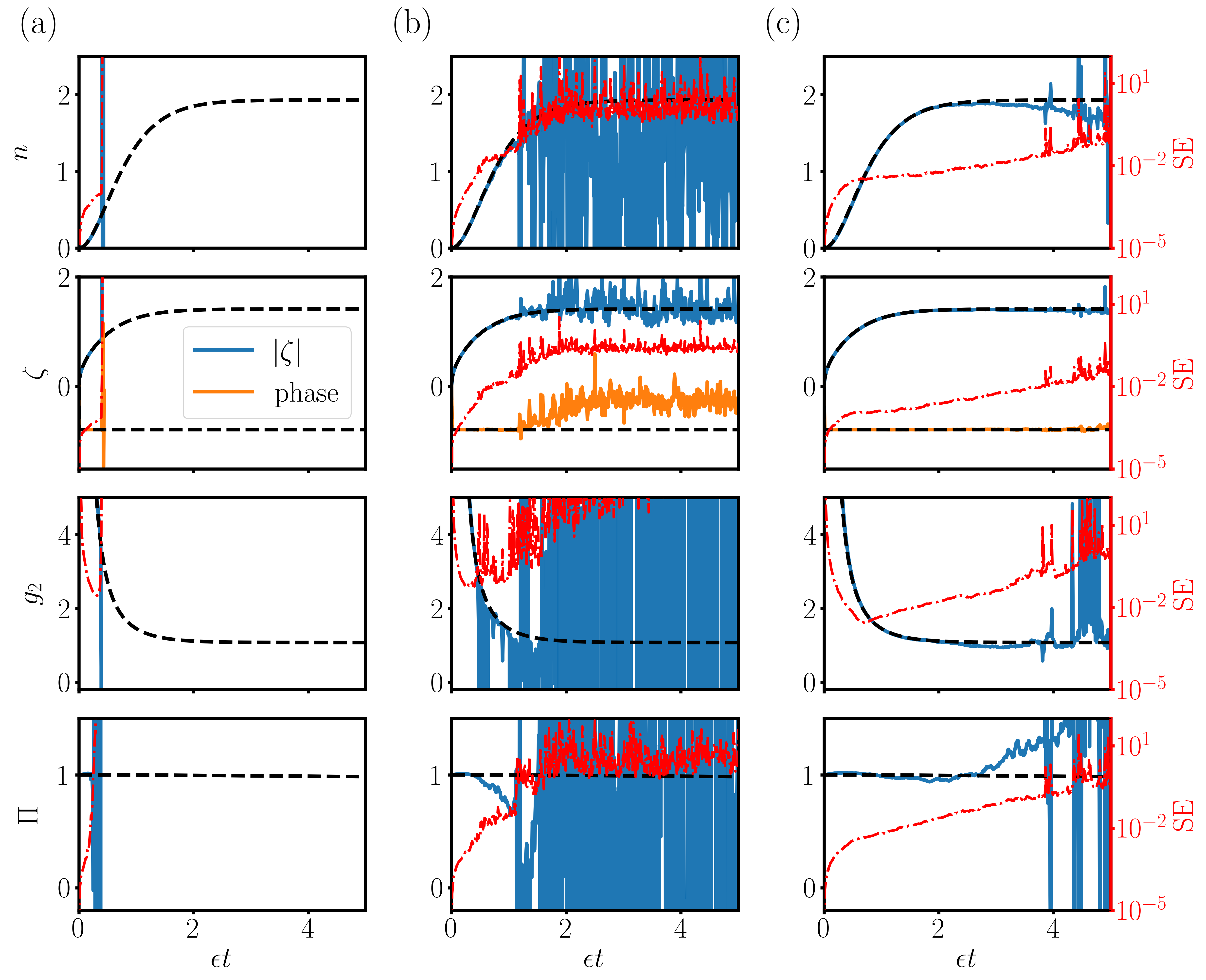}
    \caption{Comparison of observable dynamics from (a) the Choice 1 positive-\(P\) method, (b) the Choice 1 gauge-\(P\) method, and (c) the Choice 2 gauge-\(P\) method. Solid lines: stochastic results. Black dashed lines: exact master-equation results. Parameters: \(\kappa_1/\epsilon=10^{-3}\) and \(\kappa_2/\epsilon=1\).}
    \label{fig:S1}
\end{figure}

We now evaluate the effectiveness of the gauge-$P$ methods for the two diffusion matrix decompositions shown in Tab. \ref{tab:Comparison of Drift and Diffusion Matrices}. We select $\kappa_1/\epsilon = 10^{-3}$ and $\kappa_2/\epsilon = 1$, corresponding to the unstable parameter regime identified in Fig. 4(a) of the main text.

In Fig. \hyperref[fig:S1]{\ref*{fig:S1}(a)}, we examine the observable dynamics resulting from the positive-$P$ representation using Choice 1. This configuration exhibits instability similar to the original diffusion decomposition in Fig. 4(a). In Fig. \hyperref[fig:S1]{\ref*{fig:S1}(b)}, where drift gauges are included, the gauge-$P$ simulation demonstrates increased stability. Compared to both positive-$P$ schemes, this method achieves a longer duration of agreement with the master equation. However, the simulation still fails before the transient state is reached.

\begin{figure}[h!]
    \centering
    \includegraphics[width=0.6\linewidth]{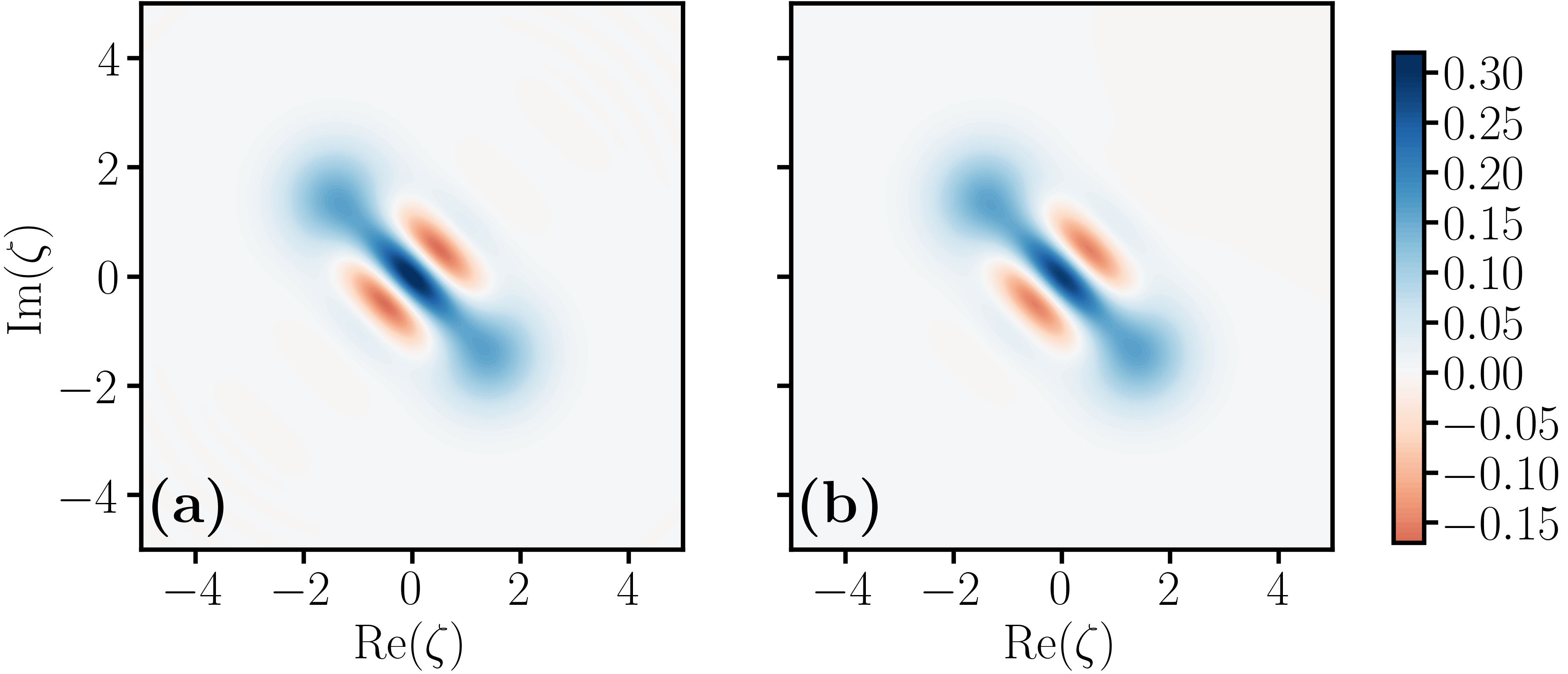}
    \caption{Wigner functions of the density matrix from (a) the master equation and (b) the gauge-\(P\) simulation. Parameters: \(\kappa_1/\epsilon=10^{-3}\), \(\kappa_2/\epsilon=1\), and \(\epsilon t=2\), corresponding to transient cat-state generation. Ensemble size: \(10^6\) trajectories.}
    \label{fig:S2}
\end{figure}

\begin{figure}
    \centering
    \includegraphics[width=0.7\linewidth]{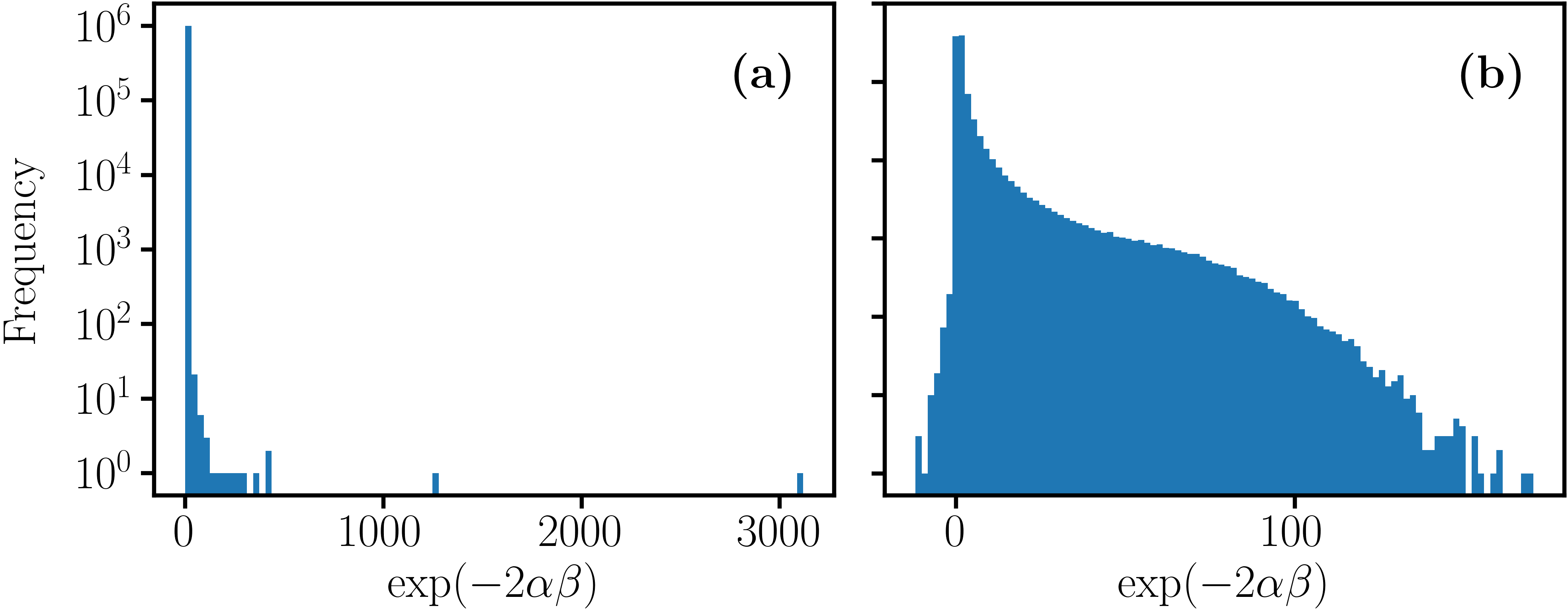}
    \caption{(a) Trajectory distribution of the parity observable \(\exp(-2\alpha\beta)\) at \(\epsilon t=3\) from the Choice 1 positive-\(P\) method. Parameters: \(\kappa_1/\epsilon=10^{-3}\) and \(\kappa_2/\epsilon=0.2\). (b) Distribution of the uncorrected parity \(\exp(-2\alpha\beta)\) at \(\epsilon t=2\) from the Choice 2 stochastic gauge. Parameters: \(\kappa_1/\epsilon=10^{-3}\) and \(\kappa_2/\epsilon=1\).}
    \label{fig:S34}
\end{figure}

Remarkably, the gauge functions in Choice 2 extend the simulation time beyond the generation of the cat state before instability eventually occurs (Fig. \hyperref[fig:S1]{\ref*{fig:S1}(c)}). Statistical errors grow slowly and smoothly before spikes appear, signaling the eventual breakdown of the simulation. During the transient dynamics, even the parity observable is modelled accurately, exhibiting strong agreement with the master equation. The Wigner function of the reconstructed density matrix displays clear interference fringes characteristic of cat states (Fig. \ref{fig:S2}). We calculate the trace of the density matrix to be $\text{Tr}(\hat{\rho}) = 0.907$, suggesting that the Choice 2 gauge functions successfully delay boundary-term errors until after the transient cat state forms. At $\epsilon t = 2$, when the cat state is produced, the long tail of the $\exp(-2\alpha\beta)$ distribution observed in the positive-$P$ simulation is eliminated (Fig. \ref{fig:S34}). This indicates that boundary errors of the second kind are also resolved on this time scale.

\begin{figure}[h!]
    \centering
    \includegraphics[width=0.3\linewidth]{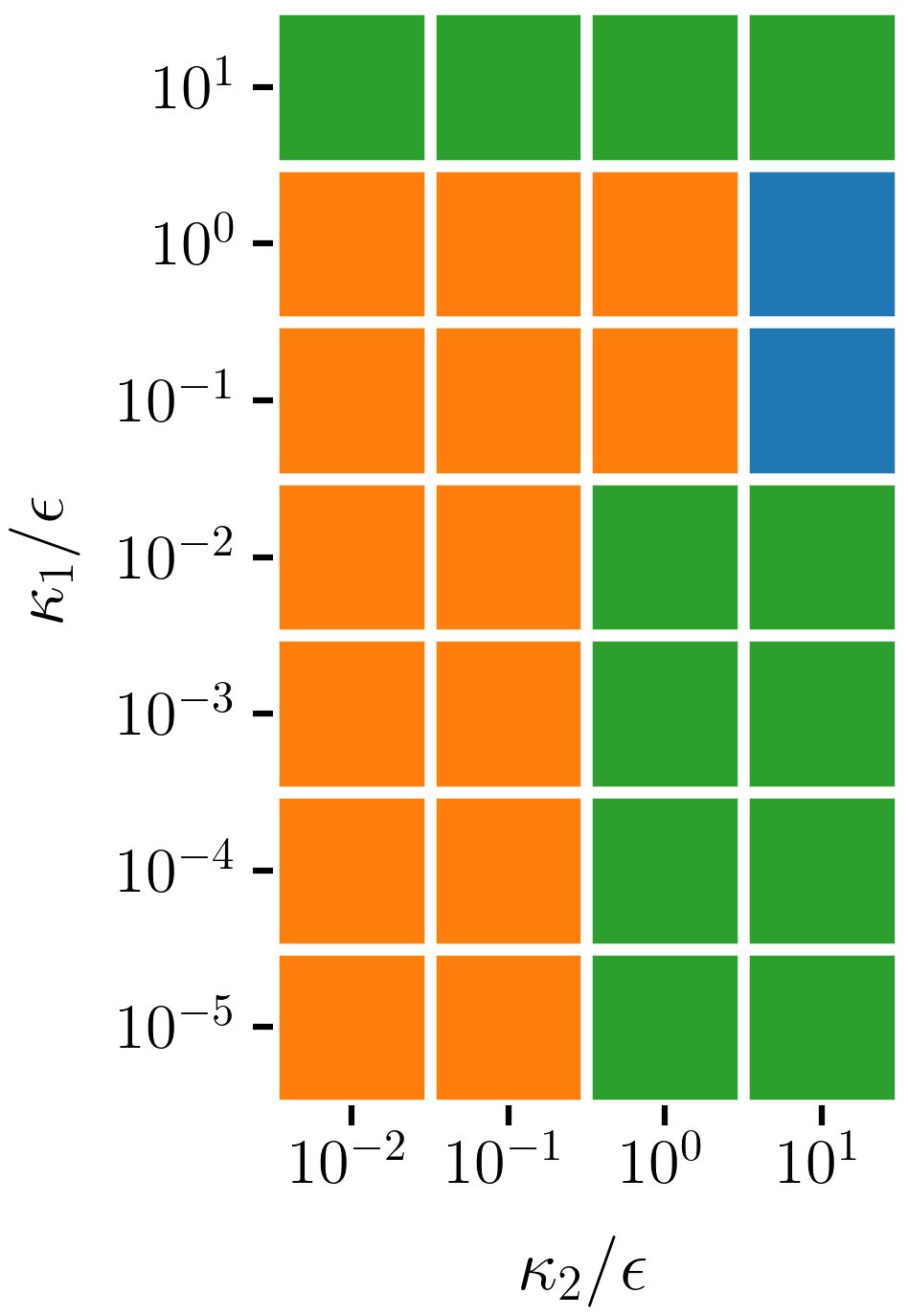}
    \caption{Effectiveness of the Choice 2 stochastic gauge function for transient dynamics across different parameter scales. Green area: excellent agreement of all observables with the master equation until the transient state. Orange area: large trajectory fluctuations (spikes) and significant signal-to-noise degradation before the transient state. Blue area: stable and accurate simulation up to the transient state, with \(g_2\) still evolving at the onset of dominant numerical noise.}
    \label{fig:S5}
\end{figure}

We also evaluate the effectiveness of the Choice 2 drift gauge across different parameter scales. For system parameters within the orange region of Fig. 3 (main text), all observables, including parity, are stabilised until the transient state is reached. However, for parameters in the blue region, the introduction of drift gauges further destabilises the trajectories. Thus, while this drift stochastic gauge stabilises trajectories in the unstable positive-$P$ regime and correctly predicts the parity observable, it does not resolve the unexpected parity decay issue in regimes where other observables are otherwise accurately predicted (Fig. \hyperref[fig:S5]{\ref*{fig:S5}}).

\subsection{2.3 Parity decay of the Cat states}

For a cat state stored in a harmonic oscillator, the dominant noise channel is single-photon loss, which causes a bit-flip by inverting the state parity. A meaningful characterisation of the system's decoherence is therefore the parity decay rate. In this subsection, we assess the utility of the positive-$P$ and gauge-$P$ methods for simulating the parity decay of cat states. The $\alpha$ and $\beta$ variables of a cat state can be sampled from the probability distribution \cite{teh2018creation}:
\begin{equation}
    P(\alpha,\beta) = \frac{1}{\mathcal{N}}[\delta_{+,+} + \delta_{-,-} + \exp(-2|\zeta|^2)(\delta_{+,-}+\delta_{-,+})],
\end{equation}
where $\mathcal{N} = \sqrt{2(1\pm\exp(-2|\zeta|^2))}$ is the normalisation factor, and $\delta_{\pm,\pm} = \delta(\alpha\pm\zeta)\delta(\beta^*\pm\zeta)$. Note that this choice of distribution is not unique and differs distinctly from the distribution obtained in the transient dynamics shown in Fig. \ref{fig:S34}.

\begin{figure}[h!]
    \centering
    \includegraphics[width=0.6\linewidth]{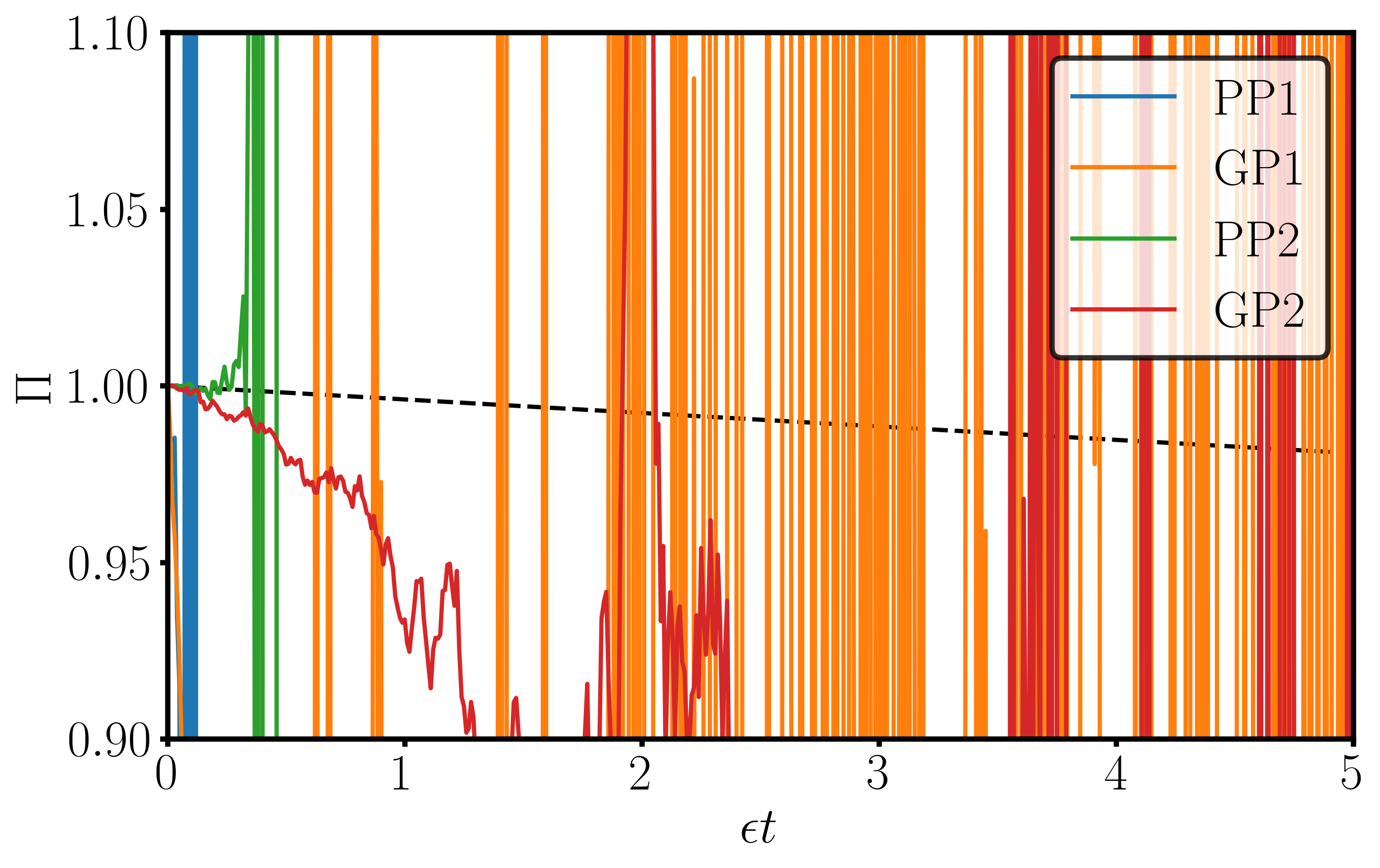}
    \caption{Parity decay dynamics of a positive cat state. Labels PP$i$ and GP$i$: positive-$P$ and gauge-$P$ schemes corresponding to Choice $i$ in Tab. \ref{tab:Comparison of Drift and Diffusion Matrices}. Black dashed line: master equation simulation.}
    \label{fig:S6}
\end{figure}

In Fig. \ref{fig:S6}, we prepare an initial positive cat state with $\kappa_1/\epsilon = 10^{-3}$ and $\kappa_2/\epsilon=1$ and analyse the parity dynamics using the four positive-$P$ and gauge-$P$ schemes listed in Tab. \ref{tab:Comparison of Drift and Diffusion Matrices}. For Choice 1, the positive-$P$ trajectories diverge in less than $\epsilon t = 0.1$. The gauge-$P$ simulation avoids immediate divergence but exhibits instability shortly thereafter. Choice 2 demonstrates similar behaviour: the gauge-$P$ simulation begins to develop instability as the positive-$P$ simulation diverges. In this instance, the simulation remains stable, albeit inaccurate, until $\epsilon t = 0.3$. These results suggest that neither the positive-$P$ nor the gauge-$P$ method can accurately capture the decay dynamics of the cat states.

\section*{Supplementary Note 3. BENCHMARKING THE MULTIMODE POSITIVE-$P$ SIMULATION}

\begin{figure*}[h!]
    \centering
    \includegraphics[width=0.9\linewidth]{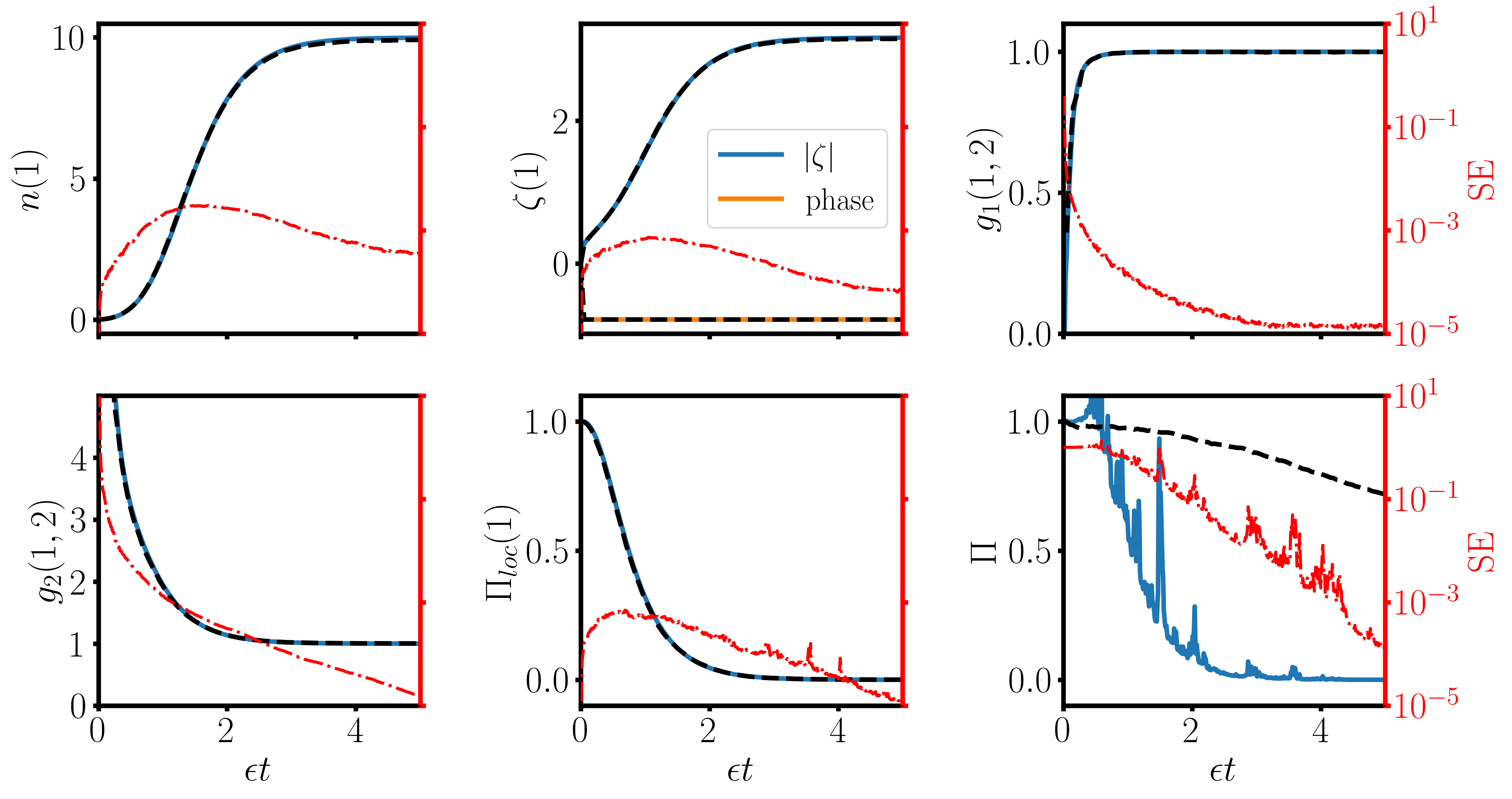}
    \caption{Comparison of observable dynamics from the positive-\(P\) representation (solid colored lines, mean in blue and statistical error in red on a logarithmic scale) and the quantum trajectory method (black dashed lines) for a small multimode system (\(N=3\)). Observables: \(n(1)\) (average photon number), \(\zeta(1)\) (coherent amplitude), \(g_{1}(1,2)\) (first-order correlation), \(g_2(1,2)\) (second-order correlation), \(\Pi_{\mathrm{loc}}(1)\) (local parity), and \(\Pi\) (global parity). Parameters: \(\kappa_1/\epsilon=0.001\), \(\kappa_2/\epsilon=0.2\), and \(\gamma/\epsilon=10\).}
    \label{fig:m1}
\end{figure*}

We assess the effectiveness of the positive-\(P\) method in estimating the transient dynamics of multimode cat-state systems. Numerical results for a small system with \(N=3\) sites near the Zeno limit, \(\gamma\gg\kappa_1,\kappa_2,\epsilon\), are presented in Fig. \ref{fig:m1}. We compare the positive-\(P\) simulation with quantum trajectory simulations performed using the mcsolve method in QuTiP \cite{johansson2012qutip,JOHANSSON20131234}.

As shown in Fig. \ref{fig:m1}, the positive-$P$ results show excellent agreement with the quantum trajectory simulations for the average photon number $n(1)$, coherent amplitude $\zeta(1)$, first-order correlation function $g^{(1)}(1,2)$, second-order correlation function $g^{(2)}(1,2)$, and local parity $\Pi_{\mathrm{loc}}(1)$.  In the single-mode case, positive-\(P\) simulations with the same parameters \(\kappa_1\) and \(\kappa_2\) exhibit an unexpected decay in parity. However, in the multimode case, the positive-\(P\) simulation of the local parity shows excellent agreement with the quantum trajectory method. This stability arises from the large amplitude of the non-local dissipator \(\gamma\), which plays a role similar to that of \(\kappa_1\) in stabilising the stochastic trajectories. For small \(\gamma\), we expect the stability behaviour to resemble that observed in the single-mode case. The global parity $\Pi$ is the only observable for which the two methods exhibit a significant discrepancy. Similar to the single-mode case, the multimode parity displays an unexpected decay before the transient state is reached. The observed spikes indicate the occurrence of boundary-term errors of the second kind \cite{deuar2005first}. This error is more severe in the multimode case because the parity operator takes the form

\[
\Pi
=
\left\langle
\exp\left(i\pi\sum_j\hat a_j^\dagger\hat a_j\right)
\right\rangle
=
\lim_{s\rightarrow\infty}
\left\langle
\exp\left(-2\sum_j\alpha_j\beta_j\right)
\right\rangle_s .
\]
The inclusion of more terms in the exponential makes the observable increasingly sensitive to fluctuations across all modes, further accelerating the accumulation of sampling errors.

While benchmarking against exact numerical methods is possible for \(N=3\), performing such comparisons for intermediate system sizes, for example \(N=5\), is computationally intractable. For a typical local truncation of 100 bosons per mode, the Hilbert-space dimension for \(N=5\) reaches \(10^{10}\), vastly exceeding the capabilities of standard quantum trajectory simulations. However, unlike approximate phase-space methods, the positive-\(P\) mapping does not introduce systematic truncation errors \cite{deuar2021fully}. Its reliability is instead limited by boundary-term and sampling errors. We therefore continuously monitor the statistical errors and the occurrence of divergent spikes as indicators of simulation instability.

We also applied the stochastic drift gauges developed for the single-mode case, as described in Supplementary Note 2, to multimode systems with both small and large values of \(\gamma\). However, we observed no significant improvement in simulation stability. In the stochastic differential equations for multimode systems, each mode acquires not only a local damping term but also the coupling terms

\[
\frac{\gamma}{2}e^{i\phi}\alpha_{j+1}
\qquad\text{and}\qquad
\frac{\gamma}{2}e^{-i\phi}\alpha_{j-1},
\]
arising from the non-local dissipation. Similar coupling terms appear in the equations for \(\beta_j\). These additional terms introduce inter-mode correlations and increase the complexity of the stochastic trajectories, reducing the effectiveness of the drift gauges developed for the single-mode case. To overcome this limitation, possible alternative approaches could  include constructing drift gauges that account for inter-mode correlations. Future studies could also explore hybrid simulation methods or emerging variational phase-space techniques, such as those employing a Variational Multi-Gaussian ansatz optimised using automatic differentiation \cite{tosca2025efficient}, which have recently shown promise in stably capturing highly non-classical states with strong Wigner negativities.

\section*{Supplementary Note 4. SECOND-ORDER SPATIAL CORRELATIONS}

\begin{figure*}[h!]
    \centering
    \includegraphics[width=0.9\linewidth]{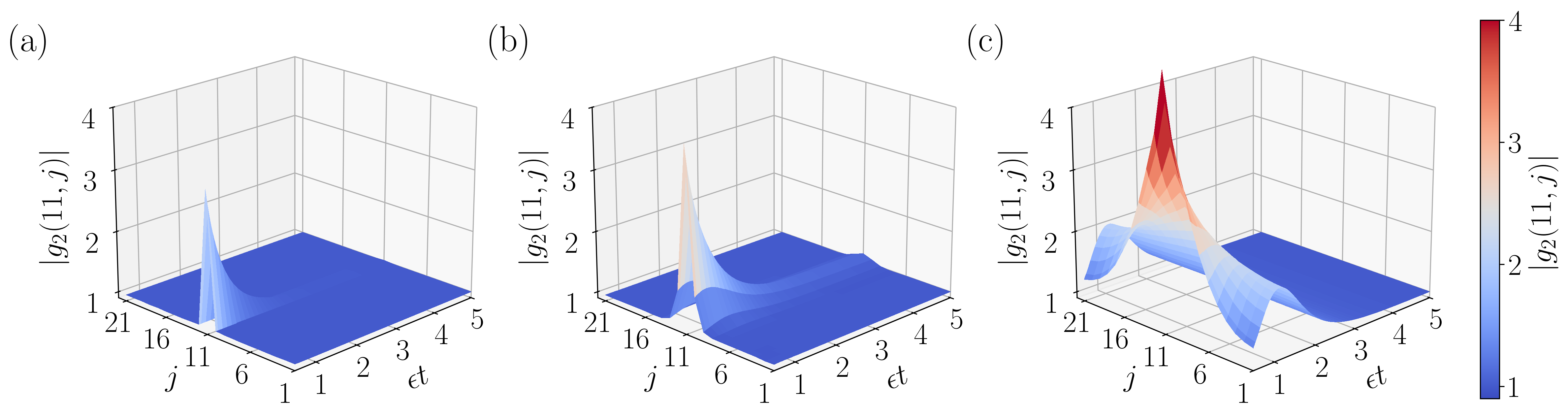}
    \caption{Time evolution of the second-order correlation function \(\lvert g_2(11,j)\rvert\) between reference site 11 and site \(j\) for \(N=21\) modes. Columns: (a) \(\gamma/\epsilon=0.2\), (b) \(\gamma/\epsilon=2\), and (c) \(\gamma/\epsilon=50\).}
    \label{fig_s7}
\end{figure*}

\begin{figure*}[h!]
    \centering
    \includegraphics[width=0.9\linewidth]{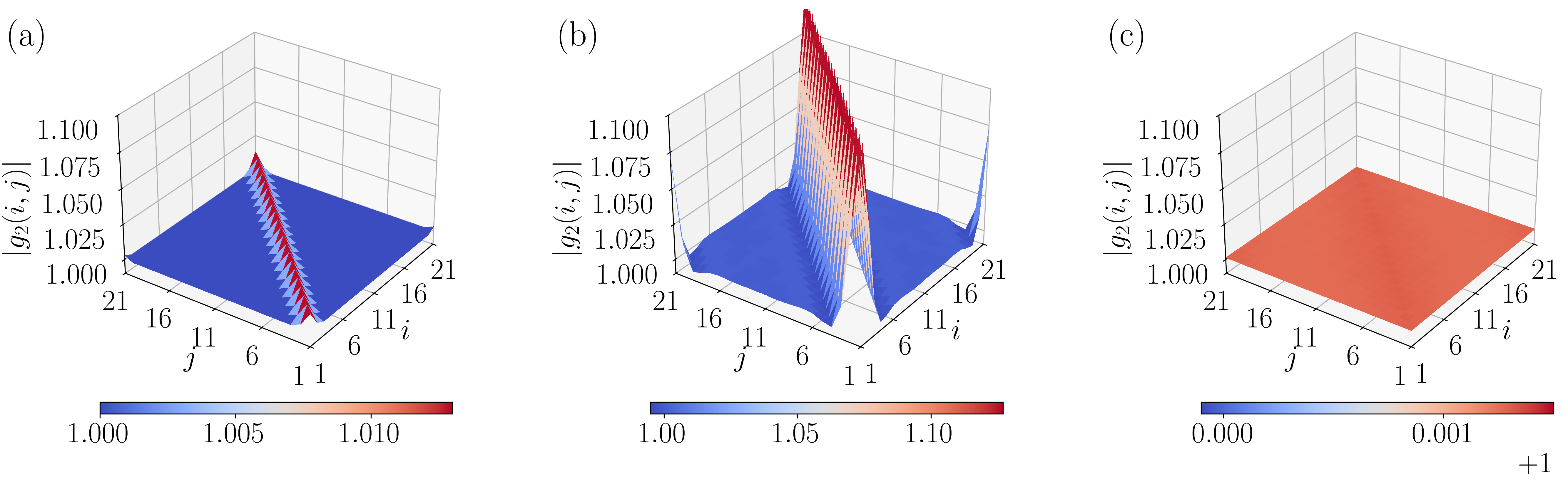}
    \caption{Snapshots of the second-order spatial correlation functions between sites \(i\) and \(j\) at \(\epsilon t=5\). Panels: (a) \(\gamma/\epsilon=0.2\), (b) \(\gamma/\epsilon=2\), and (c) \(\gamma/\epsilon=50\).}
    \label{fig_s8}
\end{figure*}

In addition to the first-order spatial correlations discussed in the main text, we examine the second-order correlation function $\lvert g^{(2)}(11,j)\rvert$ under different strengths of non-local dissipation. Fig. \ref{fig_s7} shows its time evolution between the reference site $11$ and site $j$, while Fig. \ref{fig_s8} shows the full spatial correlation function between sites $i$ and $j$ at time $\epsilon t=5$.

When weak non-local dissipation is introduced, for example $\gamma/\epsilon=0.2$, the second-order correlations remain predominantly local throughout the dynamics (Fig. \hyperref[fig_s7]{\ref*{fig_s7}(a)}). A distinct ridge appears along the diagonal of the spatial correlation function (Fig. \hyperref[fig_s8]{\ref*{fig_s8}(a)}), indicating weak local bunching, while the correlations between distinct sites remain negligible.

As the non-local dissipation is increased to $\gamma/\epsilon=2$, the second-order correlations propagate over a longer distance (Fig. \hyperref[fig_s7]{\ref*{fig_s7}(b)}). The spatial correlation function exhibits strong on-site bunching along the diagonal, with $g^{(2)}>1$, accompanied by short-range antibunching extending over approximately three sites (Fig. \hyperref[fig_s8]{\ref*{fig_s8}(b)}). These features likely arise from the competition between the coherent drive dynamics and the dissipative coupling.

In the Zeno limit, with $\gamma/\epsilon=50$, the second-order correlations rapidly spread across the full system (Fig. \hyperref[fig_s7]{\ref*{fig_s7}(c)}). At time $\epsilon t=5$, the entire $g^{(2)}(i,j)$ surface is nearly uniform and remains slightly above unity (Fig. \hyperref[fig_s8]{\ref*{fig_s8}(c)}). Together with the uniform first-order coherence discussed in the main text, this behaviour is consistent with the formation of a globally coherent multimode state.

\bibliography{citations}